\documentclass[10pt,journal]{IEEEtran}
\IEEEoverridecommandlockouts

\usepackage{cite}
\usepackage{dirtytalk}
\usepackage{amsmath}
\usepackage{amssymb}
\usepackage{amsthm}
\usepackage{amsfonts}
\usepackage{algorithm}
\usepackage{import}
\usepackage{balance}
\usepackage{braket}
\usepackage{qcircuit}
\usepackage{algpseudocode}
\usepackage{adjustbox}
\usepackage{graphicx}
\usepackage{listings}
\usepackage{caption}
\usepackage{subcaption}
\usepackage{multirow}
\usepackage{booktabs}
\usepackage[table,xcdraw]{xcolor}
\usepackage{authblk}

\def\BibTeX{{\rm B\kern-.05em{\sc i\kern-.025em b}\kern-.08em
    T\kern-.1667em\lower.7ex\hbox{E}\kern-.125emX}}

\newcommand{\goldmine}{{\scshape GoldMine}}

\begin{document}

\title{SCAR: Power \underline{S}ide-\underline{C}hannel \underline{A}nalysis at \underline{R}TL-Level \\
}


\author[1]{Amisha Srivastava}
\author[1]{Sanjay Das}
\author[1]{Navnil Choudhury}
\author[2]{Rafail Psiakis}
\author[2]{Pedro Henrique Silva}
\author[3]{Debjit Pal}
\author[1]{Kanad Basu}
\affil[1]{University of Texas at Dallas, USA} 
\affil[2]{Technology Innovation Institute – Secure Systems Research Center, P.O. box 9639}
\affil[3]{University of Illinois Chicago, USA}

\maketitle

\begin{abstract}

Power side-channel attacks exploit the dynamic power consumption of cryptographic operations to leak sensitive information of encryption hardware. Therefore, it is necessary to conduct power side-channel analysis for assessing the susceptibility of cryptographic systems and mitigating potential risks. Existing power side-channel analysis primarily focuses on post-silicon implementations, which are inflexible in addressing design flaws, leading to costly and time-consuming post-fabrication design re-spins. Hence, pre-silicon power side-channel analysis is required for early detection of vulnerabilities to improve design robustness. In this paper, we introduce SCAR, a novel pre-silicon power side-channel analysis framework based on Graph Neural Networks (GNN). SCAR converts register-transfer level (RTL)
designs of encryption hardware into control-data flow graphs and use that to detect the design modules susceptible to side-channel leakage. Furthermore, we incorporate a deep learning-based explainer in SCAR to generate quantifiable and human-accessible explanation of our detection and localization decisions. We have also developed a fortification component as
a part of SCAR that uses large-language models (LLM) to automatically generate and insert additional design code at the localized zone to shore up the side-channel leakage. When evaluated on popular encryption algorithms like AES, RSA, and PRESENT, and post-quantum cryptography algorithms like Saber and CRYSTALS-Kyber, SCAR, achieves up to 94.49\% localization accuracy, 100\% precision, and 90.48\% recall. Additionally, through explainability analysis, SCAR reduces features for GNN model training by 57\% while maintaining comparable accuracy. We believe that SCAR will transform the security-critical hardware design cycle, resulting
in faster design closure at a reduced design cost.

\end{abstract}


\section{Introduction} \label{sec:intro}


\begin{figure}[t!]
    \centering
    \includegraphics[width=\columnwidth]{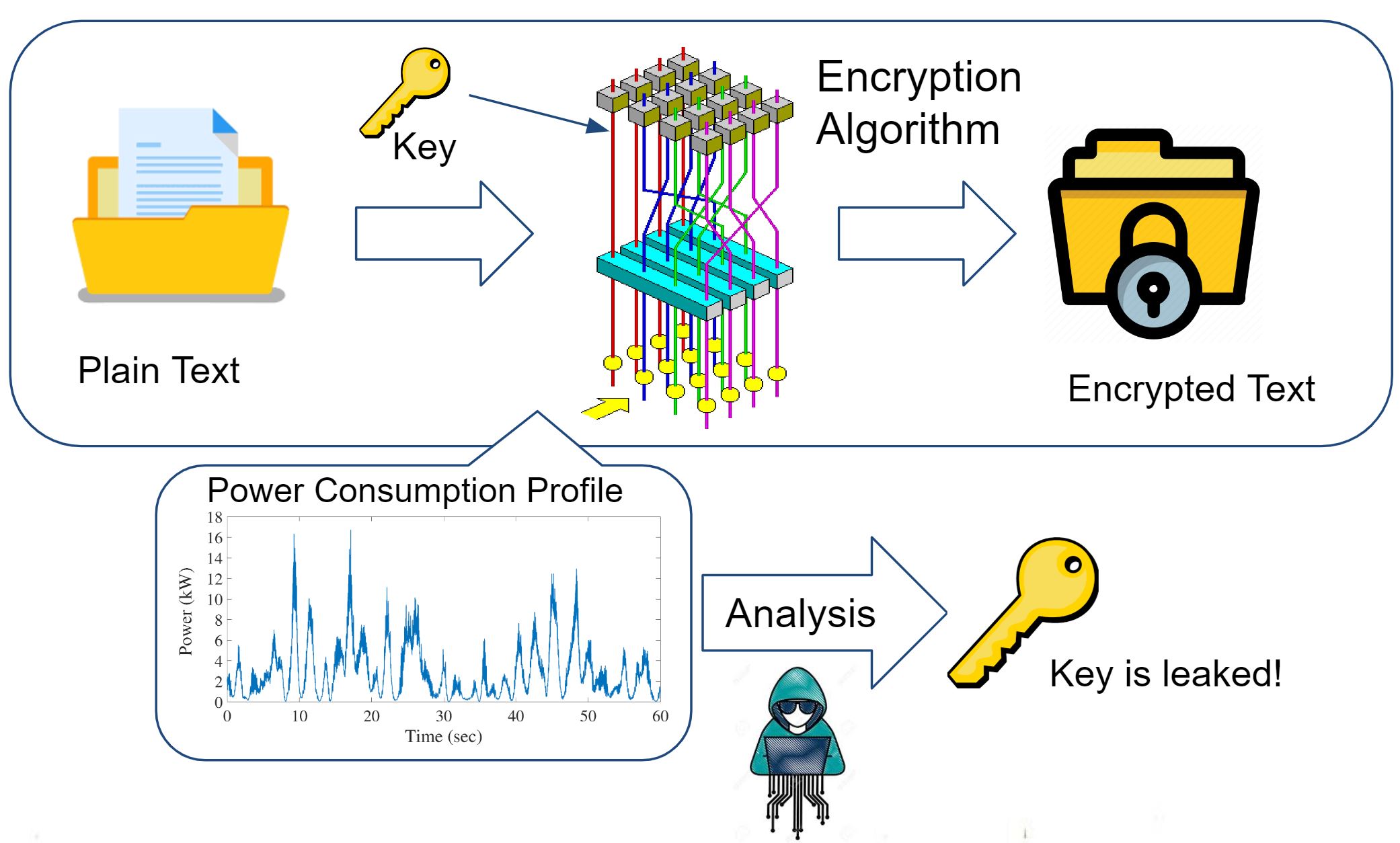}
    \vspace{-0.15in}
    \caption{Power side-channel attack flow. Attacker observes and analyzes dynamic power consumption profile of security-critical hardware devices to reconstruct secrets such as encryption keys.}
    \vspace{-0.15in}
    \label{fig:int}
\end{figure}

Power side-channel (PSC) attacks leverage variations in the dynamic power consumption exhibited during cryptographic operations to extract confidential data from the underlying encryption hardware. These attacks have become a significant concern for the security of hardware-based cryptographic systems, including smart cards, secure microcontrollers, and other embedded systems. Moreover, the attacks are effective against a broad 
range of cryptographic algorithms, including symmetric and asymmetric key algorithms, hash functions, and digital signature schemes \cite{tiri2007side}. Therefore, it is necessary to conduct PSC analysis for assessing the susceptibility of cryptographic systems to such attacks and mitigating the potential risks.
PSC analysis focuses on the power consumption exhibited by cryptographic encryption algorithms during computationally intensive operations, including bitwise operations such as \texttt{XOR}, \texttt{AND}, and \texttt{OR}~\cite{power}. By examining the power consumption pattern, 
PSC analysis can be used to infer information about such operations, leading to 
side-channel leakage. Figure \ref{fig:int} illustrates the scenario of how an attacker can obtain information about the encryption key by monitoring the power consumption of the system during encryption.

Typically, PSC 
analysis is performed by measuring the power consumption of the target device during its operation. The attacker 
uses a power probe to measure the device power consumption 
during different cryptographic operations. The analysis can be performed using various techniques such as differential power analysis (DPA) or simple power analysis (SPA) \cite{joy2011side}. One major 
drawback 
of PSC analysis at 
the post-silicon level is that it is nearly 
impossible to retrofit security measures onto existing devices leading to costly device re-spins. 
On the contrary, pre-silicon PSC analysis provides two significant advantages -- i) it 
can significantly improve the system's robustness and reduce the cost since it enables early detection of vulnerabilities in the design process and ii) it enables easy and relatively cheaper design changes than post-silicon. 
Previous research proposed pre-silicon analysis at both the layout and IC chip power modeling levels to evaluate PSC leakage, which is inflexible \cite{lin2020fast}, \cite{tsukioka2019fast}. 

At the pre-silicon level, the RTL design can be segmented into individual registers and the combinational logic which transmits the data and carries out functionalities. This abstraction can provide the designer with vital information on power consumption patterns during cryptographic operations, facilitating the identification of potential side-channel leakage regarding the encryption key. Moreover, it allows the designer to conduct necessary enhancements on the vulnerable modules with less cost since the analysis is done in the pre-silicon design phase. Existing research demonstrates PSC analysis at the RTL level by estimation of power profiles \cite{pundir2022power, 
 he2019rtl, zhang2021psc}. However, the process is computationally intensive and time-consuming due to the need for simulating power traces and using complex power estimation models, 
 particularly for complex designs. Moreover, the approach employs only modular-level analysis of the RTL designs and do not incorporate any strategies for fortification for the designs. 

In this paper, we propose a deep-learning-aided framework 
to enhance the security and robustness of cryptographic hardware against power-side channel attacks 
by preemptively identifying PSC leakage-prone locations and fortifying 
the design RTL. By transforming RTL into a control-data flow graph (CDFG), we can classify the leakage-prone module through a node-level classification task using a Graph Neural Network (GNN) model. By identifying vulnerable (\textit{i.e.}, ``leaky'') design modules, we can utilize the GNN model to capture the influence of neighboring nodes and the local 
structure of the module compared to the ``non-leaky'' modules. 
Additionally, we employ a source code analysis approach to pinpoint and isolate the vulnerable locations of the design, precisely identifying the specific lines within the RTL code that are susceptible to PSC leakage. Furthermore, we fortify  
cryptographic hardware designs against PSC attacks by augmenting vulnerable design locations with additional protective codes generated using a 
pre-trained large-language model (LLM).

Our proposed framework, SCAR, utilizes the RTL design of the encryption algorithm as the input and predicts the locations vulnerable to PSC attacks in the design. SCAR can be employed as an effective measure to prevent PSC attacks by predicting and fortifying the vulnerable locations during the design phase. The major contributions are as follows:
\begin{itemize}
  \item  We propose SCAR, a novel RTL-level PSC analysis technique which utilizes 
  CDFGs extracted from the encryption hardware RTL and deep learning on such graph 
  to identify RTL locations susceptible to PSC attacks.
  \item  SCAR includes an {\em explainer} 
  to generate quantifiable and human-accessible explanations of SCAR's 
  predictions by constructing a annotated subgraph from the CDFG annotated with 
  each feature's (information obtained from each node) importance in the prediction.

    \item We enhance the detection granularity of our framework by identifying specific lines within the vulnerable RTL modules that can cause the PSC leakage, through a source code analysis-based approach. 
      \item We utilize a LLM 
      to automatically generate and insert 
      mitigation codes to mask PSC leakage of the 
      vulnerable lines. 
      \item When compared with post synthesis results, our evaluation demonstrates a high degree of accuracy. Our identified vulnerabilities have been shown to induce fluctuations in dynamic power, further affirming the reliability of our approach.
  \item When evaluated on previously unobserved AES implementations, and encryption algorithms like RSA and PRESENT, not present in the training data, SCAR achieves up to 94.49\% accuracy, 100\% precision, and 97.88\% recall in identifying the vulnerabilities. Additionally, when applied to lattice-based Post Quantum Cryptography (PQC) algorithms Saber and CRYSTALS-Kyber, the framework achieves up to 91.84\% accuracy, 85.94\% precision, and 94.62\% recall. 
  \item Lastly, by incorporating explainability analysis, the framework also reduces the required number of features for GNN model training by 57\% while maintaining comparable accuracy, enhancing its efficiency and interpretability.
\end{itemize}

The rest of the paper is organized as follows. Section \ref{sec:background} provides the background on PSC analysis, graph neural networks and large language models. Section \ref{sec:Methodology} describes the proposed methodology. Section \ref{sec:Results} evaluates our proposed approach. Lastly, the paper is concluded in Section \ref{sec:Conclusion}.

\section{Background and Related Work}\label{sec:background}


\subsection{Power Side-Channel Analysis}

PSC analysis targets unauthorized information leakage via variation in power consumption during cryptographic algorithm execution. Techniques include differential power analysis, simple power analysis, and correlation power analysis \cite{randolph2020power}, \cite{repka2015correlation}. Prior works involve pre-silicon layout-based power-noise side-channel leakage analysis \cite{lin2020fast}. However, layout analysis can be time-consuming and less flexible. Another work focuses on power analysis in hardware designs, aiming to identify potential PSC leakage using test pattern generation \cite{zhang2021psc}.  At the post-silicon level, existing research presents a hardware implementation of a secure key exchange protocol, leveraging ring-LWE, and fortified against PSC attacks through masking and randomization \cite{aysu2018binary}. An alternative study proposes a real-time PSC attack detection technique using on-chip sensors based on a thorough analysis \cite{gattu2020power}.

\subsection{Graph Neural Network} 
Graph neural networks (GNNs) are a class of neural networks that operate on graph-structured data, such as social networks, molecular structures, and knowledge graphs \cite{zhou2020graph}. GNNs have have become popular for analyzing and processing graph-structured data in a wide range of tasks, including node classification, link prediction and graph classification \cite{gupta2021graph}. GNN models are capable of learning and encoding the local structure of the graph around each node \cite{scarselli2008graph}. The GNN-Explainer 
generates explanations for the predictions made by GNNs \cite{ying2019gnnexplainer}. It works by assigning importance scores to the input features and using them to construct a local subgraph around the output node.
Recently, GNNs have demonstrated their applicability in addressing various challenging Electronic Design Automation (EDA) problems \cite{lopera2021survey}. Existing research focuses on autonomously learning operation pattern mappings within the context of high-level synthesis \cite{ustun2020accurate}. Another study proposes the use of a GNN to extract structural characteristics from gate-level netlists and predict metrics associated with soft error propagation \cite{balakrishnan2020composing}. 

\subsection{\goldmine: A static analyzer for hardware designs}\label{sec:goldmine}

GoldMine is a multifaceted tool designed for hardware (RTL) design analysis tasks \cite{goldmine} such as assertion generation, static analysis, etc. 
\goldmine~utilizes lightweight source code analysis and machine learning (ML) to 
generate 
design assertions. \goldmine~generates high-quality assertions that capture important design behavior in an easy-to-understand manner. \goldmine~also induces a rank among generated assertions by analyzing the importance and complexity of the captured design behavior. 

\subsection{Large Language Models}
Large language models (LLMs), built upon the Transformer architecture, represent a transformative breakthrough in natural language processing \cite{chang2023survey}. These models, pretrained on vast text corpora, have shown remarkable capabilities in understanding and generating human-like text across various applications \cite{naveed2023comprehensive}, \cite{meng2023unlocking}. Recently, LLMs have been used for code generation, which is the task of automatically generating executable code from natural language specifications \cite{poesia2022synchromesh}. There have been various techniques and applications of code generation with LLMs, such as self-planning and Verilog code generation \cite{jiang2023self, thakur2023verigen}.
\section{Proposed SCAR methodology}
\label{sec:Methodology}

\begin{figure}[t!]
    \centering
    \includegraphics[width=0.70\linewidth]{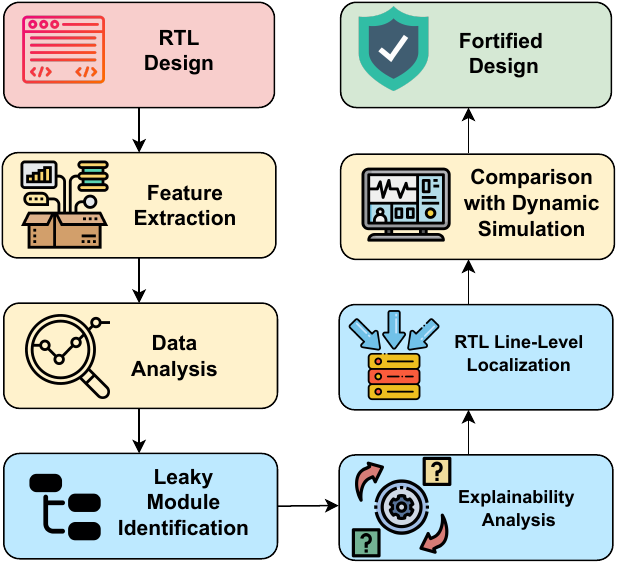}
    \caption{SCAR Framework: the red and green blocks indicate the input and output to the framework, respectively. The blue blocks highlight the three main stages which perform the PSC leakage localization. Lastly, the yellow blocks represent the intermediate steps required by the framework.} 
    \label{fig:labell}
\end{figure}

Our framework, SCAR, predicts vulnerable PSC locations in encryption hardware RTL using a GNN, where we represent RTL as a CDFG for input. Figure \ref{fig:labell} demonstrates an overview of our proposed approach. The GNN model identifies leaky nodes within the CDFG, which correspond to vulnerable modules. Furthermore, we conduct source code analysis on these identified modules, assigning leakage scores to the lines in the vulnerable modules. We then compare the high-scoring lines to leaky nodes and propose automatic masking to enhance security against PSC attacks. 

\vspace{-0.1in}
\subsection{Module Identification}
\label{sec:module}

\subsubsection{Control-Data Flow Graph Generation}

We use 
\goldmine~to generate the RTL design CDFG and to analyze it to gain design insights~\cite{goldmine}. 
Note that any other CDFG generation tools can also be used for the same. 
In order to identify the leaky modules in a design, knowledge about the power consumption of each component is essential. When an input graph, which includes variables affecting dynamic power, is provided, it can yield substantial insights from the dataset. In this context, leveraging the CDFG of the design allows us to gain valuable information about module connectivity and the data encapsulated within each module.
The nodes in the CDFG correspond to basic blocks of the RTL - sequential groups of instructions lacking branching or jumping operations. Each node gives information about the module name, line number, and the node type, which is crucial for comprehending the control and data flow within the design, aiding in the extraction of input features required for the GNN model to predict vulnerable PSC locations in encryption hardware RTL. The edges in the graph depict the control flow between these basic blocks, which could be attributed to conditional statements, loops, or case statements. Such usage facilitates the invocation of a specific module, as well as the signals employed within the corresponding location. The CDFG provides valuable insights into the interconnectivity of modules and sub-modules, as well as the precise location of variables within the design.

\subsubsection{Node Preprocessing}
In this section, we present a detailed methodology to extract pivotal node features from the CDFG of RTL designs. These features are selected based on their influence on dynamic power consumption and potential vulnerabilities to PSC attacks.  By incorporating these features into our analysis, we aim to provide a comprehensive and detailed assessment of the power characteristics and security vulnerabilities of the RTL designs.   
Following this, we highlight the features we examine for each node in the CDFG. These features are essential in generating the feature set for the GNN.


\medskip

\textbf{Number of vulnerable paths}:
If there is a strong correlation between the encryption key and a variable in the RTL, it can make the variable a potential source of leakage. This is because an attacker can observe the variable's value and use it to deduce information about the secret key. In order to prevent the disclosure of the sensitive variable (\textit{i.e.}, the key in case of encryption algorithms), we predict, for each node in the CDFG (which corresponds to a basic design block in the RTL), its connection to the encryption key. For capturing these links between variables and the sensitive variable, we utilize a variable dependency graph. A variable dependency graph visually represents the interdependencies between variables in an algorithm. In this graph, nodes represent variables in the RTL, and edges signify their dependencies, indicating how changes to one variable might impact others.

\begin{figure}[t!]
    \centering
    \includegraphics[width=\columnwidth]{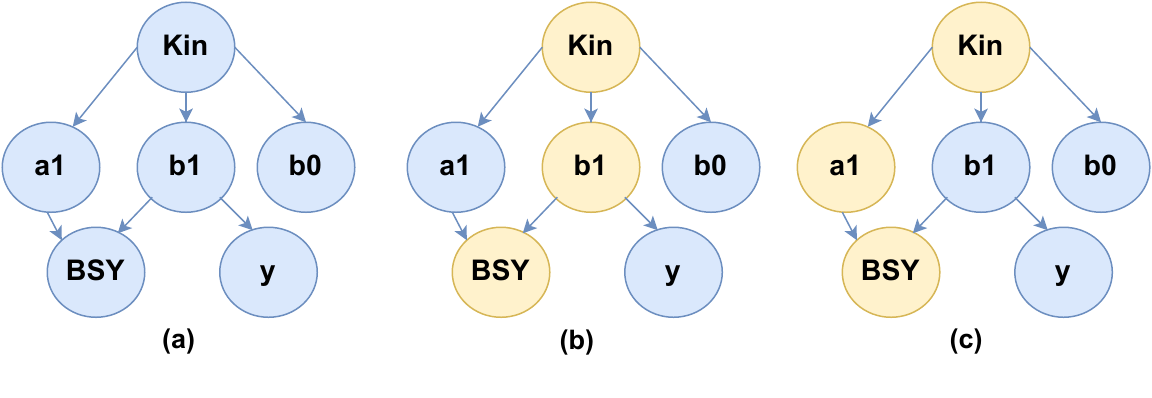}
    \caption{Variable Dependency Graph Example for a node labelled `AES\_Comp\_DEC.539:AS' in the CDFG for the RTL Source Code of AES\_Comp Design. Here, a1, b1, b0 and y represent variables in the design. Kin represents the encryption key for AES\_Comp benchmark. }
    \label{fig:vgraph}
\end{figure}

For every node in the CDFG, we can extract the variable stored at that location in the RTL design and then check the dependency of this variable on the key. 
For instance, if the extracted signal for a node in the CDFG is `BSY', and the encryption key is the variable `Kin'. These terms have corresponding nodes in the variable dependency graph. A snippet of the variable dependency graph has been shown in Figure \ref{fig:vgraph} (a). It can be seen that it contains both the encryption key (labeled as `\textit{Kin}') and the signal BSY as its nodes. In order to determine the connection of the node having its data as BSY, with the key, we perform Depth First Search (DFS) to calculate all the possible paths between this node and the key \cite{depth}.
Figures \ref{fig:vgraph} (b) and (c) show the possible paths obtained through DFS, where the yellow nodes highlight a path from `Kin' to `BSY'. All these paths are susceptible to information leakage since they have a connection to the encryption key. Moreover, the greater the number of such paths a node has, the greater the chances of it being leaky.

\medskip

\textbf{Degree of each node}:
The node degree 
provides information about the number of connections for each node. The greater the number of connections, \textit{i.e.}, the number of edges stemming in and out from a node, the higher the chances of it being connected to a leaky neighbor. A leaky neighbor corresponds to an adjacent node which is associated with a vulnerable module. 
If a node has more neighbors, it has a higher probability of being connected to a neighbor that is compromised or vulnerable to a PSC attack.

\medskip 

\textbf{Hamming Distance}:
The Hamming distance serves as a metric for comparing two binary data strings. For a precise assessment of dynamic power consumption, it is imperative to consider the number of bit-flips, given that they denote the energy expenditure of a digital circuit amidst its transitions between logic states \cite{dynamic}. The Hamming distance links dynamic power to bit-flips in RTL designs. Each time a bit in a signal toggles, it corresponds to a state change in the circuit, which is associated with power consumption from the charging and discharging of capacitors at transistor gates. Using the Hamming distance, which signifies the count of differing positions between two consecutive signal states, we bypass complex power estimation models \cite{pundir2022power} and negate the need for dynamic simulations.  Our model is trained on features tied to dynamic power, requiring only the RTL design input.

To calculate the Hamming distance, the RTL design of the encryption algorithm is simulated and results are stored in a Value Change Dump (VCD) file.  This simulation is conducted using \textit{Icarus Verilog}. The VCD file chronologically logs binary representations of all variable changes. The total Hamming distance is derived by evaluating consecutive signal states and selecting the maximum observed value.
Given a series of N signal states denoted by S, where S = ($s_{1}, s_{2}...s_{n})$, the Hamming distance between two consecutive signal states $s_{i}$ and $s_{i+1}$ is given by equation~\ref{eqn:consec_ham}.
\begin{equation}
\begin{aligned}
HD_i(sig) =  Count\_{ones}(s_{i-1} \oplus s_{i})
\end{aligned}
\label{eqn:consec_ham}
\end{equation}
Here, $HD_i(sig)$ represents the hamming distance of signal $sig$, at $i^{th}$ and $(i-1)^{th}$ state. $Count\_{ones}$ is the function that counts the number of ones in a binary string. In Equation~\ref{eqn:consec_ham}, two consecutive state values of $sig$ are XOR-ed to find differing bit values, fr
om which the number of ones give the hamming distance travelled by $sig$ while going from $(i-1)^{th}$ state to $i^{th}$ state.
\par In order to accurately capture all the bit-flips that occur during the simulation, the count value for the corresponding register is incremented each time a bit-flip occurs. This process can be illustrated using Equation~\ref{eqn:ham_calc} for the example signal, $sig$.
\begin{equation}
    \begin{aligned}
        HD_{total}(sig) = \sum_{i=1}^{N}{\lvert HD_i(sig) \rvert}
    \end{aligned}
    \label{eqn:ham_calc}
\end{equation}
Here, $HD_{total}(sig)$ represents the total hamming distance travelled by signal, $sig$ and $N$ refers to the total number of states for $sig$. This method effectively captures signal value changes and utilizes them to assess the differential power consumption between signals. Bit-flips serve as clear indicators of dynamic power consumption when comparing one signal to another.

\medskip 

\textbf{Operation Type}:
To extract information regarding dynamic power consumption, we consider logical operations (\texttt{AND}, \texttt{OR}, \texttt{XOR}, and \texttt{MUX}) that contribute to the bit flipping of signals. The arithmetic operations performed on signals also contribute to bit flips and thus can lead to power consumption. The gate operations \texttt{XOR}, \texttt{OR}, \texttt{AND} contribute to the change in the Hamming distance of the signals. Moreover, a multiplexer (\texttt{MUX}) is a form of conditional assignment to the signal and contributes to power consumption as well. For each node in the CDFG, we determine the presence of either of these operations by a binary encoded vector. 
The values of these vectors indicate the presence (value 1) or absence (value 0) of each of the following operations: \texttt{XOR}, \texttt{OR}, \texttt{AND}, \texttt{MUX}, indicating dynamic power consumption in the particular node.

\subsubsection{Graph Neural Network Model}
\label{sec:gtrain}
\begin{figure}[h!]
    \centering
    \centering
    \includegraphics[width=\columnwidth]{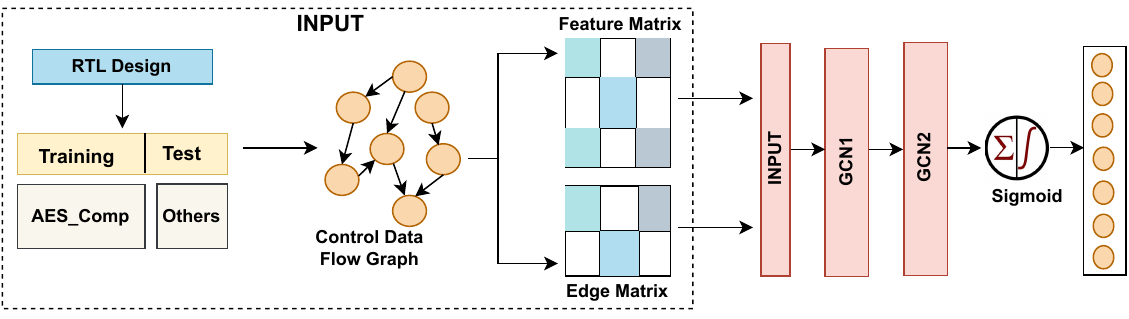}
    \vspace{-4mm}
    \caption{Graph neural network model architecture with input as RTL and output as leaky nodes. The feature matrix consists of the set of all the node features, edge matrix is a set of all the links present between the nodes. The layers GCN1 and GCN2 update the node representations by leveraging their surrounding connections within the graph.} 
     \label{fig:gmodel}
\end{figure}

Figure \ref{fig:gmodel} illustrates the process flow of our proposed GNN model.  It can be observed that only the RTL designs of the encryption algorithms are utilized as our input dataset. In order to train the GNN model, we utilize the \textit{AES\_Comp} design, which is a composite field implementation of the Advance Encryption Standard (AES) encryption algorithm \cite{aes}. The selection of AES for training the model is motivated by its established usage in the domain of PSC attack detection such as \cite{ambrose2008mute}. Moreover, \textit{AES\_Comp} was selected for training the GNN model, as it exhibited the highest number of nodes in the CDFG, compared to other benchmarks, like, \textit{AES\_PPRM1}, etc., mentioned in detail in Section \ref{sec:Results}-A. 

In this particular AES implementation, Sboxes are implemented based on the composite field. \textcolor{black}{The \textit{AES\_Comp} cipher consists of the Mixcolumns module, the Sbox module, Subbytes, Encryption and Decryption modules.} Out of these modules, it is already known from prior research that Sbox and Mixcolumns modules are vulnerable to PSC attacks \cite{vass, carlet2005highly}. 
The MixColumns operation involves multiplying each column of the state matrix by a fixed matrix of coefficients. \textcolor{black}{These multiplication operations can potentially introduce power variations that correlate with the key.} An attacker can use this variation in power consumption to deduce information about the key used in the MixColumns operation. Moreover, the Sbox operation in AES is closely related to the encryption key which makes it vulnerable to PSC attacks. 

Therefore, we label the nodes belonging to the Sbox and Mixcolumns modules as leaky for our training dataset. The rest of the nodes in CDFG will be labeled as non-leaky. 
It can be observed from Figure \ref{fig:gmodel} that the CDFG produced by utilizing the method, is used as the input graph to the GNN model. The feature set for the model comprises the node-level attributes \textendash~
node degrees, Hamming distance, vulnerable paths, operations \texttt{AND}, \texttt{OR}, \texttt{XOR}, and \texttt{MUX}. 

In the architecture of the GNN model, the input layer comprises seven neurons, 
one neuron for each feature. 
There are two graph convolutional layers in this architecture, namely, GCN1 and GCN2, as shown in Figure \ref{fig:gmodel}. Each of these layers is composed of two hidden layers. Each graph convolutional layer takes a tuple of three elements as input: node representations, edges, and edge weights, each of which represent the node features, the edges between nodes, and the weights of each edge, respectively. The weights for each edge are assigned as one in this model, since all connections are considered equally important. The output layer of the network is a fully connected feedforward neural network layer. This layer takes the final node representations from the GNN and applies a linear transformation, followed by a non-linear activation function to produce the final output. This output consists of a single neuron with a sigmoid activation function, indicating that the problem is binary classification, with the output representing the probability of the input graph belonging to the leaky or non-leaky class. The GNN model generates predictions for each node in the test dataset, categorizing them into leaky or non-leaky class.

\subsection{Explainability of the GNN model}
\label{sec:explain}

In the domain of detecting vulnerabilities within encryption designs, ensuring accurate predictions is of paramount importance. However, it is equally critical to ascertain the underlying reasons for these predictions since they provide insights into the model's decision-making mechanism. It facilitates the identification of the most relevant features contributing to the model's decisions. Consequently, features with negligible contributions can be efficiently identified and eliminated, leading to a reduction in computational overhead. 

In this section, our primary objective is to provide meaningful interpretations for the predictions generated by our GNN model for leaky module detection. To achieve this goal effectively, we have leveraged a perturbation-based explanation technique, specifically GNN-Explainer, due to its suitability for addressing our specific problem \cite{ying2019gnnexplainer}. Importantly, we emphasize that while we employ GNN-Explainer in our analysis, our framework remains versatile and adaptable to other explanation techniques as well. This explainability analysis is performed in order to construct a subgraph $G_s \subseteq G$ (the CDFG) and associated node features $X_s = \{x_i|v_i \subseteq G_s\}$, that are critical in terms of their contribution to the node's classification.
Once the GNN model is trained, it is used to generate forecasts for sets of nodes that were not encountered during the training process. 
 To elucidate a specific prediction, the GNN-Explainer is provided with the model, a target node, and the entire nodeset with its features and edges. The Explainer subsequently generates feature importance scores and a corresponding subgraph, delineating whether a node is classified as leaky or non-leaky. Individual node explanations contribute to creating a global feature importance map for the entire GNN model. By assessing and ranking these individual scores, we discern the top-ranked features' significance and their impact on the model's performance. This information is fed back for improving the proposed GNN-based vulnerable module detection methodology.



\subsection{Vulnerability Localization}
\label{sec:localize}
 \vspace{-0.05in}
\begin{figure}[b!]
    \centering
    \includegraphics[width=0.70\columnwidth]{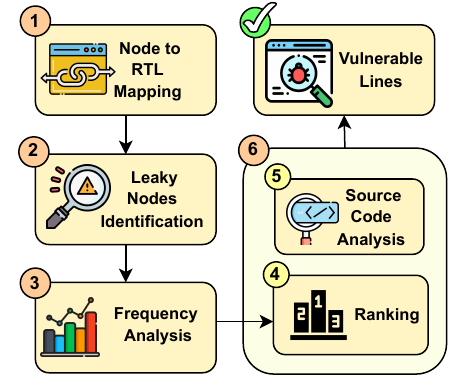}
    \caption{ Each block refers to a step in the vulnerability localization methodology. It can be observed that the process begins by associating the GNN nodes with the lines in the RTL source code, culminating in the identification of the vulnerability-prone lines.}
    \label{fig:vulne}
\end{figure}

Following the determination of leaky modules evaluated by our GNN model in Section \ref{sec:module}, we shift our focus towards localizing the precise lines of code within the encryption modules which induces leakiness.
The localization process in question takes place at the RTL source code level. This process assumes a pivotal role in the precise determination of the specific locations where leakage phenomena manifest within the previously identified modules, resulting in vulnerabilities. Consequently, in this section, we establish a methodology for identifying such vulnerabilities at the RTL source-code level. This process of correlating the leaky nodes identified by the GNN model to specific lines in the RTL code for encryption modules involves several key steps, which are subsequently outlined, as shown in Figure \ref{fig:vulne}.


 \subsubsection{Node to RTL Line Mapping} 
 \label{sec:nodeRTL}

Each node within the GNN is assigned a unique identifier that connects the GNN nodes to specific lines of code within RTL modules, including sub-modules. This one-to-one mapping establishes a direct link between GNN nodes and the corresponding lines of RTL code, enabling precise pinpointing of vulnerable code segments. This mapping acts as a crucial bridge between the GNN's predictions of vulnerable modules, and the exact lines of code evoking these vulnerabilities in the identified modules. As a result, it significantly enhances the framework's capability to detect and address potential PSC vulnerabilities at the RTL line-level within the encryption algorithm.


\subsubsection{ Leaky Node Identification}

The GNN model is trained to categorize nodes as either `leaky' or `non-leaky', based on previously computed features. Once  the `leaky' nodes are identified, as described in Section~\ref{sec:module}, we proceed to collect the lines stored in these nodes, using the procedure mentioned in Section~\ref{sec:nodeRTL}. The rationale underpinning the aggregation of nodes in the `leaky' category resides in their inherent capability to establish direct associations with the susceptible RTL lines, thereby facilitating the process of localization of potential PSC leakage sources with a high degree of precision.


\subsubsection{Frequency Analysis}

Following the aggregation of the `leaky' nodes in the previous step, an analysis is conducted to determine the frequency of association between each line of code and the `leaky' nodes. For each line of code identified as vulnerable, we count the number of `leaky' nodes that are pointing to it. This step provides a frequency distribution of leakiness for each line, thus delineating the lines of code that exhibit a higher incidence of being directed to by `leaky' nodes. Consequently, these lines are accorded a elevated vulnerability status. 

\subsubsection{Vulnerability Ranking}
 \label{sec:ranking}
By utilizing the frequency distribution, we rank each line of the RTL source code. The ranking is necessary to determine the lines contributing the greatest to PSC leakage, thereby eliciting the highest degree of vulnerability. The ranking is determined by the number of leaky nodes that are correlated to each line. Lines with higher frequencies of leaky nodes are assigned higher ranks, implying they are more likely to be sources of PSC leakage and should be prioritized during implementing mitigation strategies. This is because these lines are more frequently called upon by the encryption algorithm to perform highly power consuming calculations.

\subsubsection{Source Code Analysis}
 \label{sec:source}
In order to validate the rankings obtained using the GNN predictions, we need to compare them with the quantitative evaluation of the RTL code's susceptibility to PSC leakage. In order to perform this evaluation, we assign a leakage score to each line in the RTL code of the identified vulnerable modules. This steps aids in assessing the impact of each line on the PSC leakage. The leakage score is calculated by summing the individual scores of the metrics considered for the analysis of lines. The following metrics (conditional dependency, bit manipulation, function calls, dependency depth) are utilized in our method for identifying leaky lines of code in cryptographic modules: 


\begin{itemize}
    \item \underline{Conditional Dependency}:
The presence of conditional statements like if-else and case can introduce different signal transitions and higher switching activity. This varying behavior increases the potential for side-channel attacks that exploit timing or power consumption variations. 

\definecolor{codegreen}{rgb}{0,0.6,0}
\definecolor{codegray}{rgb}{0.5,0.5,0.5}
\definecolor{codepurple}{rgb}{0.58,0,0.82}
\definecolor{backcolour}{rgb}{0.98,0.98,0.95}

\lstdefinestyle{mystyle}{
  backgroundcolor=\color{backcolour}, commentstyle=\color{codegreen},
  keywordstyle=\color{magenta},
  numberstyle=\tiny\color{codegray},
  stringstyle=\color{codepurple},
  basicstyle=\ttfamily\footnotesize,
  breakatwhitespace=false,         
  breaklines=true,                 
  captionpos=b,                    
  keepspaces=true,                 
  numbers=left,                    
  numbersep=5pt,                  
  showspaces=false,                
  showstringspaces=false,
  showtabs=false,                  
  tabsize=2,
  escapechar=/n,
  xleftmargin = 10pt
}
\lstset{style=mystyle}
\lstset{style=mystyle}
\begin{lstlisting}[language=Verilog] 
 if (secret_key == input) then output = 1; else output = 0;
\end{lstlisting}
In this example, the code line contains an if-else statement that depends on secret data (secret\_key). Therefore, the presence of one condition checking statement will contribute a score of one to the total leakage score.

\item \underline{Bit Manipulation}:
Operations involving bit manipulation, such as bitwise AND, OR, XOR, shifts and rotations, can lead to increased dynamic power consumption. The more bit manipulation operations in a line, the higher the potential for leakiness.

\begin{lstlisting}[language=Verilog] 
assign new_val = (old_val << 4) ^ (old_val >> 2);
\end{lstlisting}
In this example, the line contains two shifts and an XOR operation, thereby making the contribution due to the operations as three in the total leakage score.

\item \underline{Function Calls}:
The presence of function calls within a line could potentially introduce additional PSC vulnerabilities, depending on the nature of the functions being called. The leakiness of the function itself would contribute to the leakiness score of the line making the call.  For example, if a function accesses a lookup table, the actual computational cost or operations occur during the function's execution (when it is called), not during its definition. Therefore, the function calling line inherits the underlying operational and data-handling properties of the function.
\lstset{style=mystyle}
\begin{lstlisting}[language=Verilog] 
output = Encrypt(input, key);
\end{lstlisting}
The example shows that the line contains a function call to Encrypt, thereby, adding a score of one to the total leakage score.

\item \underline{Dependency Depth}:
Each line's dependency on other lines contributes to its `depth'. The greater the number of lines a given line depends on, either through direct or indirect data flow, the more complex the behavior and thus potentially higher the leakiness. 
\lstset{style=mystyle}
\begin{lstlisting}[language=Verilog] 
assign z = x + y; assign w = z * 2;
\end{lstlisting}
In this case, the second assignment line (assign w = z * 2) is dependent on the first (assign z = x + y), thereby having a dependency depth of 1.
\end{itemize}

Now, let us consider the following RTL code in order to determine the value of each metric and calculate the total leakiness score:

\definecolor{codegreen}{rgb}{0,0.6,0}
\definecolor{codegray}{rgb}{0.5,0.5,0.5}
\definecolor{codepurple}{rgb}{0.58,0,0.82}
\definecolor{backcolour}{rgb}{0.98,0.98,0.95}

\lstdefinestyle{mystyle}{
  backgroundcolor=\color{backcolour}, commentstyle=\color{codegreen},
  keywordstyle=\color{magenta},
  numberstyle=\tiny\color{codegray},
  stringstyle=\color{codepurple},
  basicstyle=\ttfamily\footnotesize,
  breakatwhitespace=false,         
  breaklines=true,                 
  captionpos=b,                    
  keepspaces=true,                 
  numbers=left,                    
  numbersep=5pt,                  
  showspaces=false,                
  showstringspaces=false,
  showtabs=false,                  
  tabsize=2,
  escapechar=/n,
  xleftmargin = 10pt
}
\lstset{style=mystyle}
\begin{lstlisting}[language=Verilog] 
if (private_data == flag) begin
    assign bit_transform = (data_val >> 3) | (data_val << 5);
    result = Decrypt(bit_transform, secret);
    assign sum = a + b;
    assign product = sum * c;
end

\end{lstlisting}

In this sample, a conditional check against \textit{private\_data} incurs a Conditional Dependency score of one. Bit manipulations on \textit{data\_val} result in a Bit Manipulation score of three due to two shifts and an OR operation. The Decrypt function contributes a score of one for Function Calls. Finally, a Dependency Depth score of one is noted, as the product relies on the sum derived from \textit{a} and \textit{b}. We assume that all the metrics have equal weight, the calculated overall leakage score for this block would be six, the sum of all these individual scores.

\subsubsection{Correlation with Source Code Analysis}
After obtaining the ranking from the GNN, as mentioned in Section \ref{sec:ranking}, we compare this information with the leakiness scores obtained from source code analysis in Section \ref{sec:source}, as it aides in validating the predictions made by the GNN. For instance, if a line in the leaky MixColumns module had a high leakiness score in source analysis and also ranked high utilizing the results furnished by the GNN model, then its vulnerability is strongly corroborated. Therefore, by correlating the leakiness scores of all the lines with their respective rankings, we obtain the vulnerable lines from the encryption designs.

Overall, the significance of this localization process indicates that lines with more leaky nodes rank higher in vulnerability, as they play a crucial role in encryption computations. Lines with a greater number of leaky nodes are more frequently implicated in the calculations integral to the encryption algorithm. This elevated usage and importance makes such lines more susceptible to PSC attacks. By evaluating lines against their leaky node associations, we discern the most vulnerable segments of the RTL code. Detailed results of this method have been discussed in Section \ref{sec:rtlline}.

\subsection{Fortification}
\label{sec:fortify}
Masking is essential in encryption algorithms to prevent adversaries from deducing sensitive cryptographic keys or data. This is because analyzing observable PSC information can inadvertently leak confidential information during encryption operations. Our framework leverages a pre-trained LLM, 
capable of understanding and manipulating complex code structures, in order to automatically perform masking on the vulnerable locations. In this section, we highlight the dataset utilized for training the model and its outcomes.

A commonly used approach to protect implementations of AES against power side channel attacks is Boolean masking where the goal is to randomize all intermediate results that occur during the computation of the algorithm \cite{blomer2004provably}. In this masking technique, cryptographic computations are fortified against PSC attacks by introducing a layer of obfuscation to the sensitive operations. At the beginning of each operation, a random masking value is chosen. This mask remains constant for the entirety of the operation to maintain consistency. Each sensitive operand is concealed by XORing it with this random mask, effectively masking the original data during the computational process. After executing the cryptographic operations on these masked operands, the resultant output undergoes an ``unmasking" process, achieved by again XORing it with the same random mask. This ensures that the final outcome matches what would have been achieved without the masking process, while the intermediate steps remain shielded from potential exploits. By segmenting the computation into multiple masked stages, this technique bolsters security by making it challenging for adversaries to discern the original data from observable operations.

The following example highlights our Boolean masking process.

\definecolor{codegreen}{rgb}{0,0.6,0}
\definecolor{codegray}{rgb}{0.5,0.5,0.5}
\definecolor{codepurple}{rgb}{0.58,0,0.82}
\definecolor{backcolour}{rgb}{0.98,0.98,0.95}

\lstdefinestyle{mystyle}{
  backgroundcolor=\color{backcolour}, commentstyle=\color{codegreen},
  keywordstyle=\color{magenta},
  numberstyle=\tiny\color{codegray},
  stringstyle=\color{codepurple},
  basicstyle=\ttfamily\footnotesize,
  breakatwhitespace=false,         
  breaklines=true,                 
  captionpos=b,                    
  keepspaces=true,                 
  numbers=left,                    
  numbersep=5pt,                  
  showspaces=false,                
  showstringspaces=false,
  showtabs=false,                  
  tabsize=2,
  escapechar=/n,
  xleftmargin = 10pt
}
\lstset{style=mystyle}
\lstset{style=mystyle}
\begin{lstlisting}[language=Verilog] 
\\Original expression
assign y = x^z;

\\masked counterpart
assign mask = $random;
assign y = (x^mask)^(z^mask);
\end{lstlisting}
As seen in this example, each sensitive operation is randomized by splitting it into multiple masked operations, thereby obfuscating the data being processed. First, a random mask is generated and the operands \textit{x} and \textit{z} are individually XORed with this random mask, effectively masking or ``hiding" their original values.  This way of masking is effective and generalizable since it does not depend on the previous instances of the variables present in the line and only on the current time operation being performed.

\subsubsection{Large Language Model}

Cryptographic algorithms, despite their mathematical robustness, often become susceptible to vulnerabilities when translated into real-world designs. Manually identifying and rectifying these vulnerabilities, especially in complex RTL designs, is not only time-consuming but also prone to human error. The specialized knowledge required to identify and rectify the vulnerabilities is both scarce and expensive. LLMs, with their ability to process vast amounts of information and recognize intricate patterns, present an efficient alternative. Their automated nature ensures consistency, reducing human error and offering a scalable solution to bolster cryptographic security

Our methodology employs a Large Language Model (LLM), specifically a pre-trained Falcon-7B model, to enhance the security of cryptographic algorithms. Falcon LLM is a generative open-source LLM developed by the Technology Innovation Institute (TII). Falcon-7B is a 7B parameters causal decoder-only model built by TII and trained on 1,500B tokens of RefinedWeb data enhanced with curated corpora \cite{penedo2023refinedweb}. We begin with a curated dataset containing pre-identified vulnerable lines from various RTL designs. The model is then fine-tuned on this dataset to learn how to transform these vulnerable lines into their secure, masked counterparts. The training process involves encoding both the masked and original lines of code, aiming to teach the model how to predict the masked versions of lines effectively. Upon successful validation, the model is integrated into the test benchmarks. During deployment, it takes the pre-identified vulnerable lines as input and autonomously replaces them with secure, masked lines generated during training. This automated methodology substantially improves the resilience of cryptographic designs, eliminating the need for manual code revisions, thereby providing a robust and efficient security solution.


\subsubsection{Dataset Generation}
Our fortification strategy involves the generation of a dataset that will be required to train the LLM model. The dataset contains vulnerable lines from various AES benchmarks utilized from Trusthub \cite{trust}. It also consists of synthetic examples manually generated to copy the behaviour of the lines that can pose potential risks to PSC attacks. Therefore, the synthetically generated data replicates operations and statements found in the Trusthub designs, ensuring its relevance and resemblance to real-world designs. We use synthetic data due to the limited number of designs in Trusthub that are contributing to PSC attacks, namely, AES-T300, AES-T1300, AES-T1400, AES-T1500, AES-T600, AES-T2000 and AES-T2100 \cite{kumar2017analysis}. After the selection of susceptible lines from the benchmarks, we proceed to generate their masked counterparts. 
\section{Experimental Results} \label{sec:Results}
In this section, we will describe the evaluation of our proposed technique, SCAR, using popular encryption algorithms.



\subsection{Experimental Setup}


\textbf{Model Architecture and Hyperparameters}: As mentioned in Section \ref{sec:gtrain}, the GNN architecture consists of an input layer of seven neurons, which accounts for each unique feature in the data. Furthermore, there are two pivotal graph convolutional layers, termed GCN1 and GCN2, as shown in Figure \ref{fig:gmodel}. The output is a fully connected feedforward neural network layer and undertakes a linear transformation on the derived node representations. The output is a single neuron facilitated by the sigmoid activation function. Furthermore, hyperparameter tuning was performed to achieve optimal model performance on the test dataset. The learning rate was set to 
 0.01, in order to guide the optimization process. Data was processed in batches of 20, optimizing resource utilization. A dropout rate of 0.3 was employed to enhance robustness, ensuring a random 30\% of nodes were excluded during each iteration for balanced learning. The training of the model was conducted over 32 epochs, ensuring thorough iteration over the dataset. 

\textbf{Description of AES benchmarks}:
We conduct an evaluation of our proposed approach, SCAR, using side-channel vulnerabilities in three distinct implementations of the Advanced Encryption Standard (AES), namely \textit{AES\_TBL}, \textit{AES\_PPRM1}, and \textit{AES\_PPRM3} \cite{aes}. The generated CDFGs for each of these benchmarks consist of 63, 62 and 423 nodes, respectively, providing comprehensive coverage for our analysis. Details about training the model using the AES\_Comp benchmark, whose CDFG consists of 703 nodes, have already been mentioned in Section \ref{sec:gtrain}. By testing the model on a diverse set of AES implementations, it becomes exposed to a wider range of architectural and implementation variations. This helps us verify whether the model can effectively identify vulnerable modules regardless of the specific implementation details or architectural variances. 

\textbf{Description of other benchmarks}: To demonstrate the feasibility of our proposed approach, we evaluated our model on encryption algorithms unknown to the framework, like RSA and PRESENT, which were not present in the training set, utilizing the \textit{RSA1024\_RAM} and \textit{PRESENT} benchmarks, respectively.
The CDFGs generated for these benchmarks comprised 436 and 158 nodes, respectively. Please note that our model was trained using only the \textit{AES\_Comp} benchmark. In the RSA algorithm, the Modular Exponentiation operation is susceptible to PSC leakage \cite{rsa}. Modular exponentiation is a critical operation in the RSA encryption algorithm, used to compute the ciphertext from the plaintext and the public key. The operation involves computing the remainder of a large number raised to a \textit{power}, divided by another large number. The \textit{power} is determined by the public key, and the remainder is the ciphertext. This operation is vulnerable to PSC attacks since it involves multiple conditional branches, which depend on the bits of the key. As a result, the \textit{power} operation may consume different amounts of power depending on the specific value of the key.

For the PRESENT encryption scheme, it is known that the Sbox is most vulnerable to PSC attacks for secret key leakage \cite{present}. During the encryption process, the plaintext undergoes Sboxes to produce ciphertext bits, with the output determined by plaintext and key bits. Observing power consumption patterns, an attacker can deduce the input and consequently, the key value.


Furthermore, we evaluate our framework on two lattice-based Post Quantum Cryptography (PQC) algorithms: Saber and CRYSTALS-Kyber. Both of these algorithms have achieved finalist status in the PQC Standardization process conducted by NIST. Polynomial multiplication, a crucial operation in these algorithms, is susceptible to PSC attacks due to its computationally-intensive nature. An attacker can exploit this to produce variable power consumption patterns and can potentially uncover secret operands \cite{mujdei2022side}. Consequently, modules executing these multiplications in RTL designs risk PSC leakage \cite{park2022pqc}.

Next, we will describe each of the open source benchmark designs in detail utilized for the evaluation of our framework.


\begin{itemize}
  \item \textbf{AES\_TBL}\cite{aes}: This implementation of AES utilizes look-up tables to compute the substitution box (\textit{Sbox}) output. The \textit{AES\_TBL} cipher comprises the following modules: an \textit{Encryption} module, a \textit{SubBytes} module, four \textit{MixColumns} modules, and four \textit{Sbox} modules.

  \item \textbf{AES\_PPRM1}\cite{aes}: 
  In this particular implementation of AES, the Sboxes are implemented based on Arithmetic Normal Form (ANF), which is a mathematical expression used to represent Boolean functions. The benchmark consists of four \textit{Mixcolumns} modules, four \textit{Sbox} modules, a \textit{Subbytes} module and an \textit{Encryption} module. 

  \item \textbf{AES\_PPRM3}\cite{aes}: With respect to this AES implementation, the Sboxes are implemented based on the 3-stage Perfect Power Reduction Method (PPRM), which is a mathematical technique used to minimize the implementation complexity of Boolean functions. The benchmark consists of four \textit{Mixcolumns} modules, four \textit{Sbox} modules, a \textit{Subbytes} module and an \textit{Encryption} module. 

  \item \textbf{RSA1024\_RAM}\cite{aes}: This implementation of RSA encryption algorithm involves several interconnected modules, including \textit{Memory}, 
\textit{ SequencerBlock, ModExpSequencer, MontMultSequencer, MontRedcSequencer, InvNSequencer, CpSequencer, LoopController, MemoryAddressController, MultiplicationBlock}, and \textit{ArithCore}. These modules work together to perform operations such as modular exponentiation, multiplication, and address management, among others. 

  \item \textbf{PRESENT}\cite{present}:This implementation of the PRESENT cipher utilizes three key modules: the \textit{Pbox} (permutation box) module, the \textit{Sbox} (substitution box) module, and the \textit{Sboxkey} module. These modules collaborate to ensure robust and efficient encryption. The \textit{Pbox} handles data permutation, the \textit{Sbox} manages data substitution, and the \textit{Sboxkey} operates as a specialized Sbox, generating a lookup table using a secret key.

  \item \textbf{SABER} \cite{imran2021design}: The Saber implementation executes 256-bit polynomial multiplication in parallel, using power-of-two moduli to bypass modular reduction. The primary module, \textit{wrapper\_top}, embeds \textit{ComputeCore3} and \textit{shift\_registers}. Within \textit{ComputeCore3} are essential SABER components like \textit{AddRound, AddPack,} and \textit{BS2POLVEC}. The design also integrates \textit{BinomialSampler, Vector Polynomial Multiplier, Unpack, CopyWords}, and for polynomial multiplication, it employs \textit{VectorMul\_wrapper, poly\_mul256\_parallel\_in2, parallel\_Mults1}, and \textit{small\_alu1}.

  \item \textbf{KYBER} \cite{yaman2021hardware}: 
The benchmark offers a hardware implementation for polynomial multiplication in the CRYSTALS-Kyber PQC scheme. The main module, \textit{KyberHPM1PE\_top}, incorporates \textit{KyberHPM1PE}, memory modules (\textit{BRAM} and \textit{BROM}), an \textit{addressgenerator} for memory, and \textit{butterfly} for polynomial tasks. Stages are managed by \textit{dt0} to \textit{dt3}. Arithmetic operations utilize \textit{intmul} for multiplication and other modules like \textit{modadd}, \textit{modred}, \textit{modmul}, \textit{modsub}. Lastly, \textit{shiftreg} handles data shifting.

 
\end{itemize}





\subsection{Evaluation of the Identified Vulnerable Modules} 

\begin{table}[h]
\centering
\caption{Benchmark evaluation results on predicting the vulnerable modules of the encryption algorithms by the proposed framework. }
\label{table:res}
\newcolumntype{M}[1]{>{\centering\arraybackslash}m{#1}}
\renewcommand{\arraystretch}{1.2}
\begin{tabular}{|M{1.68cm}|M{2cm}|M{1cm}|M{1cm}|M{1cm}|}
\hline
\textbf{Benchmark} & \textbf{Encryption Algorithm} & \textbf{Accuracy (\%)} & \textbf{Precision (\%)} & \textbf{Recall (\%)} \\ \hline
AES\_PPRM1   & \multirow{3}{*}{AES} & 94.49 & 100 & 90.48 \\ \cline{1-1} \cline{3-5} 
AES\_PPRM3   &                      & 93.38 & 99.22 & 93.85 \\ \cline{1-1} \cline{3-5} 
AES\_TBL     &                      & 90.48 & 97.36   & 88.90 \\ \hline
RSA1024\_RAM & RSA                  & 90.62 & 91.82 & 97.88 \\ \hline
PRESENT      & PRESENT              & 90.50 & 86.36 & 96.20 \\ \hline
SABER        & SABER                & 91.84 &      85.94 &     94.62   \\ \hline
KYBER        &KYBER                 &       88.89 &  95.38    &91.20        \\ \hline
\end{tabular}%
\end{table}

We perform RTL line-level analysis on the benchmarks to identify the leaky locations in their respective RTL-level encryption designs. First, the GNN model performs node-level binary classification to label each node. Afterwards, the nodes belonging to the ``leaky'' class are identified as vulnerable. 
The GNN model was first trained on the \textit{AES\_Comp} benchmark and furnished 98\% accuracy, due to reasons mentioned in Section \ref{sec:Methodology}. We assign training labels to each of the nodes in the generated CDFG for the benchmark. We assign the label ``leaky'' to all the nodes belonging to the Sbox Modules (SB0, SB1, SB2, SB3) and MixColumns Modules (MX0, MX1, MX2, MX3). 

On evaluation, the GNN model predicts the vulnerable modules in the RTL designs for AES, RSA and PRESENT benchmarks and obtains promising results. The performance for each of the benchmarks is outlined in Table \ref{table:res}. Column 1 of the table provides details about the benchmarks used. Column 2 refers to the encryption algorithm for each benchmark. The third column and fourth column refer to the accuracy and precision achieved for the benchmarks, respectively. Finally, the fifth column corresponds to the recall obtained by the GNN model. The proposed approach, when evaluated on previously unseen AES implementations which were not included in the training data, furnishes up to 94.49\% accuracy, 100\%  precision, and 97.88\% recall. Moreover, the generalizability of the approach is evident when it is applied to unfamiliar encryption algorithms, such as RSA, PRESENT, SABER and KYBER which possess distinct designs and functionalities compared to AES. Despite not being observed during the training phase, our framework successfully identified the vulnerabilities in these designs' leaky modules. This detection was achieved by leveraging the GNN model's ability to learn the distinctive feature patterns associated with the leaky nodes. In these cases, the model achieves promising results with an accuracy of up to 90.62\%, recall of 91.82\%, and precision of 97.88\%. This indicates the potential adaptability of our approach to diverse encryption techniques, encompassing various architectures and implementations. For each of the benchmarks, the proposed GNN model successfully predicts the ``leaky'' node and thus, their corresponding leaky modules. 

\begin{table*}[h!]

\caption{Summary of RTL Line-level Localization and Fortification Results. The table highlights the lines with the highest PSC leakage in the vulnerable modules of each benchmark. To achieve the fortified counterpart, Boolean masking has been applied on the vulnerable line using LLM.}
\label{tab:summary}
\resizebox{\textwidth}{!}{%
\begin{tabular}{@{}m{4cm}m{4cm}m{3cm}m{7cm}m{8cm}@{}}
\toprule
  \multicolumn{1}{c}{\textbf{Benchmark}} &
  \multicolumn{1}{c}{\textbf{Vulnerable Module}} &
  \multicolumn{1}{c}{\textbf{Functional Statement No.}} &
  \multicolumn{1}{c}{\textbf{Vulnerable Line}} &
  \multicolumn{1}{c}{\textbf{Masked Implementation}} \\ \toprule
 \multirow{2}{*}{AES\_PPRM1} & SBOX & 1 & assign y[0] =  \newline
            x[0] \& x[2] \& x[3] \& x[4] \& x[5] \& x[6] \& x[7] \newline
	      $\Hat{}$ x[0] \& x[1] \& x[2] \& x[4] \& x[5] \& x[6] \& x[7] \newline
	      // ... \newline
            $\Hat{}$ x[0] \newline
            $\Hat{}$ 1'b1;
 & assign y[0] =  \newline
            x[0] \& x[2] \& x[3] \& x[4] \& x[5] \& x[6] \& x[7] \newline
	      $\Hat{}$ x[0] \& x[1] \& x[2] \& x[4] \& x[5] \& x[6] \& x[7] \newline
	      // ... \newline
            $\Hat{}$ x[0] \newline
            $\Hat{}$ 1'b1 \newline
            $\Hat{}$ mask;
\\ \cline{2-5}

 & Mixcolumns & 3 & assign y = {a2[7] $\Hat{}$ b1[7] $\Hat{}$ b3[6], a2[6] $\Hat{}$ b1[6] $\Hat{}$ b3[5], \newline
              // .. \newline
              a1[1] $\Hat{}$ b3[1] $\Hat{}$ b0[0] $\Hat{}$ b0[7], a1[0] $\Hat{}$ b3[0] $\Hat{}$ b0[7]}; \newline
              
 & assign y = {a2[7] $\Hat{}$ b1[7] $\Hat{}$ b3[6] $\Hat{}$ mask, a2[6] $\Hat{}$ b1[6] $\Hat{}$ b3[5] $\Hat{}$ mask, \newline
              // .. \newline
              a1[1] $\Hat{}$ b3[1] $\Hat{}$ b0[0] $\Hat{}$ b0[7] $\Hat{}$ mask, a1[0] $\Hat{}$ b3[0] $\Hat{}$ b0[7] $\Hat{}$ mask}; \newline
\\ \midrule

\multirow{2}{*}{AES\_PPRM3} & SBOX & 14 & assign y[4] = (d[3] \& a[1]) $\Hat{}$ (d[1] \& a[3]) $\Hat{}$ (a[0] \& d[0]) \newline
                $\Hat{}$ (b[3] \& d[3]) \newline
                // ... \newline
            $\Hat{}$ (b[0] \& d[0]); \newline
 & assign y[4] = (d[3] \& a[1]) $\Hat{}$ (d[1] \& a[3]) $\Hat{}$ (a[0] \& d[0]) \newline
                $\Hat{}$ (b[3] \& d[3]) \newline
                // ... \newline
    $\Hat{}$ (b[0] \& d[0]) $\Hat{}$ mask; \newline
\\ \cline{2-5}

 & Mixcolumns & 5 & assign y = {a2[7] $\Hat{}$ b1[7] $\Hat{}$ b3[6], a2[6] $\Hat{}$ b1[6] $\Hat{}$ b3[5], \newline
              // .. \newline
              a1[1] $\Hat{}$ b3[1] $\Hat{}$ b0[0] $\Hat{}$ b0[7], a1[0] $\Hat{}$ b3[0] $\Hat{}$ b0[7]}; \newline
              
 & assign y = {a2[7] $\Hat{}$ b1[7] $\Hat{}$ b3[6] $\Hat{}$ mask, a2[6] $\Hat{}$ b1[6] $\Hat{}$ b3[5] $\Hat{}$ mask, \newline
              // .. \newline
              a1[1] $\Hat{}$ b3[1] $\Hat{}$ b0[0] $\Hat{}$ b0[7] $\Hat{}$ mask, a1[0] $\Hat{}$ b3[0] $\Hat{}$ b0[7] $\Hat{}$ mask}; \newline
\\ \midrule

\multirow{2}{*}{AES\_TBL} &  SBOX &  1 & case (x) \newline
    0: S = 99; \newline
    1: S = 124; \newline
    2: S = 119; \newline
    // ...          \newline
    255: S = 22; \newline
    endcase
& case (x) \newline
    0: S = 99 $\Hat{}$ mask; \newline
    1: S = 124 $\Hat{}$ mask; \newline
    2: S = 119 $\Hat{}$ mask; \newline
    // ...          \newline
    255: S = 22 $\Hat{}$ mask; \newline
endcase

\\ \cline{2-5}
           
& MixColumns & 5 & assign y = {a2[7]  $\Hat{}$ b1[7]  $\Hat{}$ b3[6], \newline        
            a2[6] $\Hat{}$ b1[6]  $\Hat{}$ b3[5], \newline
            a2[4]  $\Hat{}$ b1[4] $\Hat{}$ b3[3] $\Hat{}$ b3[7],\newline
            //... \newline} & 
            assign y = {a2[7]  $\Hat{}$ b1[7]  $\Hat{}$ b3[6]  $\Hat{}$ mask, \newline        
            a2[6] $\Hat{}$ b1[6]  $\Hat{}$ b3[5] $\Hat{}$ mask, \newline
            a2[4]  $\Hat{}$ b1[4] $\Hat{}$ b3[3] $\Hat{}$ b3[7] $\Hat{}$ mask,\newline
            //... \newline}
          
\\ \midrule
\multirow{2}{*}{RSA1024\_RAM} & ModExpSequencer & 1 & if (Msb == 1) pc $<$= {pc[10:0],1'b0};  & if (Msb $\Hat{}$ mask == 1) pc $<$= {pc[10:0], 1'b0}; \\ \cline{2-5}
  & MultiplicationBlock & 5 & assign regy\_in = \newline assign exor[i] = d[i] $\Hat{}$ Inv;  & assign exor[i] = d[i] $\Hat{}$ mask $\Hat{}$ Inv; \\ \midrule

 PRESENT & SBOX & 1  & always @(idat) \newline 
           case (idat) \newline 
                4'h0 : odat = 4'hC; \newline 
                4'h1 : odat = 4'h5; \newline 
                // ... \newline 
                endcase  & always @(idat $\Hat{}$ mask) \newline 
    case (idat $\Hat{}$ mask) \newline 
        4'h0 : odat = odat $\Hat{}$ mask $\Hat{}$ 4'hC; \newline 
        4'h1 : odat = odat $\Hat{}$ mask $\Hat{}$ 4'h5; \newline 
        // ...  \newline 
    endcase\\ \midrule
\multirow{3}{*}{SABER} & poly\_mul256\_parallel\_in2 & 6 & secret $<$= \{s\_vec\_64, secret[1023:64]\}; & secret $<$= \{s\_vec\_64 $\Hat{}$ mask, secret[1023:64] $\Hat{}$ mask\};
\\ \cline{2-5}
 & poly\_mul256\_parallel\_in2 & 6 & secret $<$= \{secret[1019:0], secret[1023:1020] $\Hat{}$ 4'b1000\}; & secret $<$= \{secret[1019:0] $\Hat{}$ mask, (secret[1023:1020] $\Hat{}$ 4'b1000) $\Hat{}$ mask\};
\\ \cline{2-5}
 & small\_alu1 & 2 & wire [12:0] result = s[3] ? Ri - a\_mul\_s : Ri + a\_mul\_s; & wire [12:0] result = (s[3] ? (Ri  $\Hat{}$ mask - a\_mul\_s  $\Hat{}$ mask) : (Ri  $\Hat{}$ mask + a\_mul\_s  $\Hat{}$ mask));
\\ \midrule

KYBER & intmul & 1  & 
always @* P\_DSP = A*B;\newline 
assign P = P\_DSP;
 & always @* P\_DSP = (A $\Hat{}$ mask)*(B $\Hat{}$ mask);\newline 
assign P = P\_DSP;
\\ \bottomrule
\end{tabular}%
}
\end{table*}

For the \textit{AES\_PPRM1}, \textit{AES\_PPRM3} and \textit{AES\_TBL} benchmarks, the four Sbox modules (SB0, SB1, SB2, SB3) and MixColumns modules (MX0, MX1, MX2, MX3) were identified as ``leaky''. The lines highlighted in the second column of Table \ref{tab:summary} are responsible for their leakiness.
Similarly, for the \textit{RSA1024\_RAM} benchmark, the \textit{ModExpSequencer} and \textit{MultiplicationBlock} were identified as the vulnerable modules, which are prone to PSC leakage. The \textit{ModExpSequencer} module is responsible for controlling the sequence of operations in modular exponentiation, which involves a large number of modular multiplications and modular reductions. Similarly, the \textit{MultiplicationBlock} module was responsible for executing the modular multiplication operation. As mentioned previously, in RSA modular exponentiation is a critical operation susceptible to power-side channel attacks.
For the \textit{PRESENT} benchmark, we observed that the module vulnerable to PSC attacks is \textit{Sbox}. As mentioned previously, the \textit{Sbox} module of PRESENT cipher, in particular, is susceptible to PSC attacks because it involves a large number of computations that can cause variations in power consumption. By carefully measuring the power consumption of the device during the encryption or decryption process, an attacker can potentially obtain information about the secret key used in \textit{Sbox}. 

In the case of the \textit{SABER} benchmark, it was observed that multiplication-related modules are especially vulnerable to PSC attacks. Specifically, \textit{poly\_mul256\_parallel\_in2} module, which handles 256-bit polynomial multiplication and the \textit{small\_alu1} module which performs the arithmetic and accumulation operations for polynomial, contributed to the vulnerability of the design. Due to the computational intensity of these multiplication operations, there are notable variations in power consumption. This makes them susceptible targets for attackers aiming to exploit PSC weaknesses to obtain unauthorized information.
For KYBER, the \textit{intmul} module was identified as leaky. The \textit{intmul} module performs integer multiplication, which is utilized as a sub-module in modular multiplication, which is often targeted in power analysis attacks. It is also utilized in the \textit{butterfly} module which performs the main finite field arithmetic operations in the design. These operations are sensitive as their power consumption can leak information about the data being processed. 

Our experiment hypothesized that certain modules within cryptographic designs, due to their computational nature, are inherently more susceptible to PSC attacks. Our findings crystallized this in two key insights. First, it became apparent that modules handling complex mathematical tasks were especially susceptible to PSC leakage. Secondly, it underscored the importance of early detection and mitigation of the vulnerabilities.

\subsection{RTL Line-Level Analysis and Fortification Results}
\label{sec:rtlline}
This is followed by the RTL line-level analysis of the identified vulnerable modules. Table~\ref{tab:summary} presents a concise overview of identified susceptible lines in the benchmark modules. The first column specifies the benchmark name, followed by the second column indicating the module predicted as vulnerable by the GNN. The third column highlights the specific line number within the module susceptible to PSC attacks, referred to as functional statement number, since it comprises of blocks performing assignments, conditional statements as well as functions. The fourth column presents the content of the identified vulnerable line within the functional statement block. Lastly, the fifth column displays the masked version of the vulnerable line, processed by the LLM. For brevity purposes, we have not displayed the full lines from the RTL source codes of the designs. In summary, this table offers insights into vulnerabilities within various designs while also presenting the corresponding protected implementations achieved through the LLM model. The masked implementations were achieved by applying Boolean masking to the sensitive variables in the identified vulnerable lines. Consequently, our framework not only identifies but also effectively mitigates the vulnerabilities, streamlining the path to a more robust design.


\subsection{Analyzing GNN predictions}
\begin{figure}%
    \centering
    \subfloat[\centering ]{{\includegraphics[height=3.6cm]{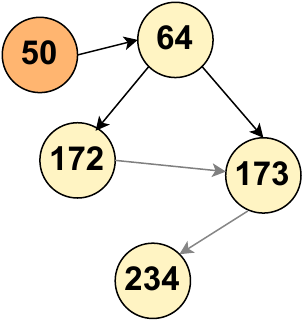} }}%
    \qquad
     \subfloat[\centering ]{{\includegraphics[height=3.8cm]{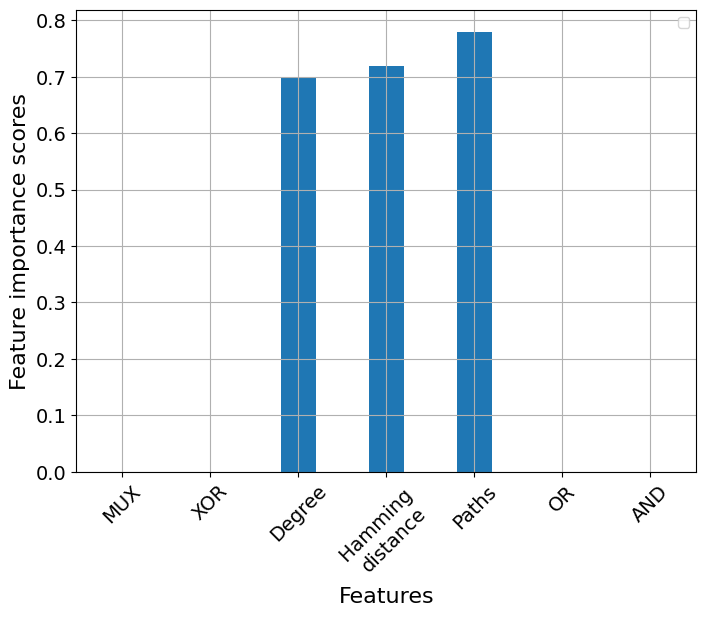} }}%
    \caption{(a) Subgraph for the instance-wise explanation, which illustrates a portion of the GNN's internal connectivity, which is relevant for understanding how the model made a specific prediction for a given instance or input. (b) Feature importance scores for a sample predicted by the GNN model, which provides information about the importance of different features in the GNN's decision-making process for that particular prediction. }
    \label{fig:example}%
\end{figure}

As explained in Section \ref{sec:explain}, we incorporate explainability analysis into the decision processes of the trained GNN model. To this end, we employ a perturbation-based explanation technique GNN-Explainer for furnishing explanations for model predictions \cite{ying2019gnnexplainer}. The GNN-Explainer is able to detect condensed subgraph structures and node characteristics that are extremely important for individual predictions of the model.
For instance, Figure \ref{fig:example} (a) depicts the generated subgraph for a random node from the PRESENT benchmark, and the corresponding feature importance scores are demonstrated in Figure \ref{fig:example} (b). In the context of instance-wise explanations, the subgraph focuses on the neighborhood and connections surrounding the node. It captures the subset of nodes and edges that directly influence the prediction of the 
GNN model. The features \say{Degree}, \say{Hamming distance}, and \say{Paths} are observed to have the most contribution with the highest score, signifying their criticality for the prediction.

\begin{figure}[t!]
    \centering
    \includegraphics[width=0.85\linewidth]{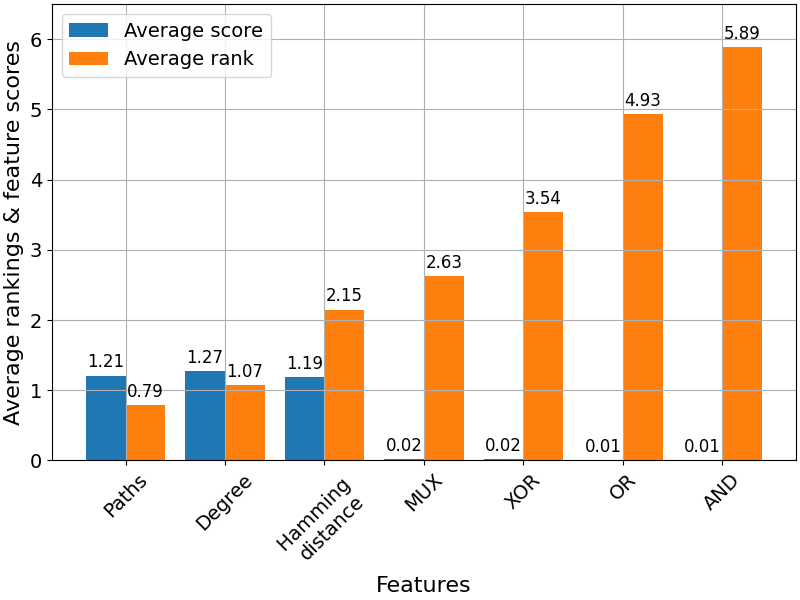}
    \caption{Graph showing the aggregated feature ranking of instance-wise explanations, showcasing the key features' significance.}

    \label{fig:exp_global}
\end{figure}

\begin{figure}[t!]
    \centering
     \includegraphics[width=\linewidth]{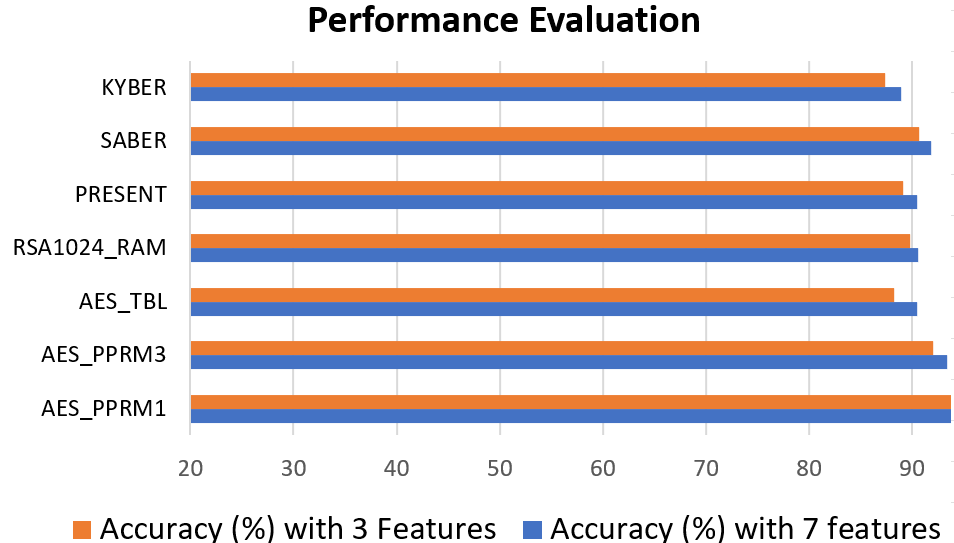}
    \caption{Graph showing the performance of all the benchmarks, when the GNN is trained with all 7 features vs only with the top-3 features.}

    \label{fig:all_Exp}
\end{figure}

In order to analyze the significance of the various node features for the model behavior as a whole, we aggregate and average the feature importance scores and their overall rankings for all the node explanations. As illustrated in Figure \ref{fig:exp_global}, it can be observed that the features  \say{Paths}, \say{Degree}, and \say{Hamming distance} are ranked as the most essential features. These top three features have similar average feature scores as well. On the other hand, the remaining four features are shown to have negligible significance in terms of lower average feature scores and thus higher average rankings.
We exploit these insights provided by the GNN-Explainer to select the most important node features and modify the dataset by removing features with negligible contribution to the model predictions, which in turn simplifies the model as well as reduces the computation overhead without compromising the model accuracy. The performance of the model when trained with only 3 features, for all the benchmarks has been shown in Figure \ref{fig:all_Exp}. It can be observed that in the case of the \textit{AES\_PPRM1} benchmark, there is a minute change in the accuracy of the GNN model. It decreases slightly from 94.49\% (when trained with all seven node features) to 93.81\% when trained only with the top-3 essential features. These three significant features when selected from the set of seven features, result in 57\% reduction of the feature set.
This demonstrates that model explanation techniques not only provide insights into node classification but also provide valuable feedback, that are useful for optimizing the model.



\subsection{Comparison with Post Synthesis Experiments}

\begin{figure*}[t!]
     \centering
     \begin{subfigure}[b]{0.18\textwidth}
         \centering
         \includegraphics[width=\textwidth]{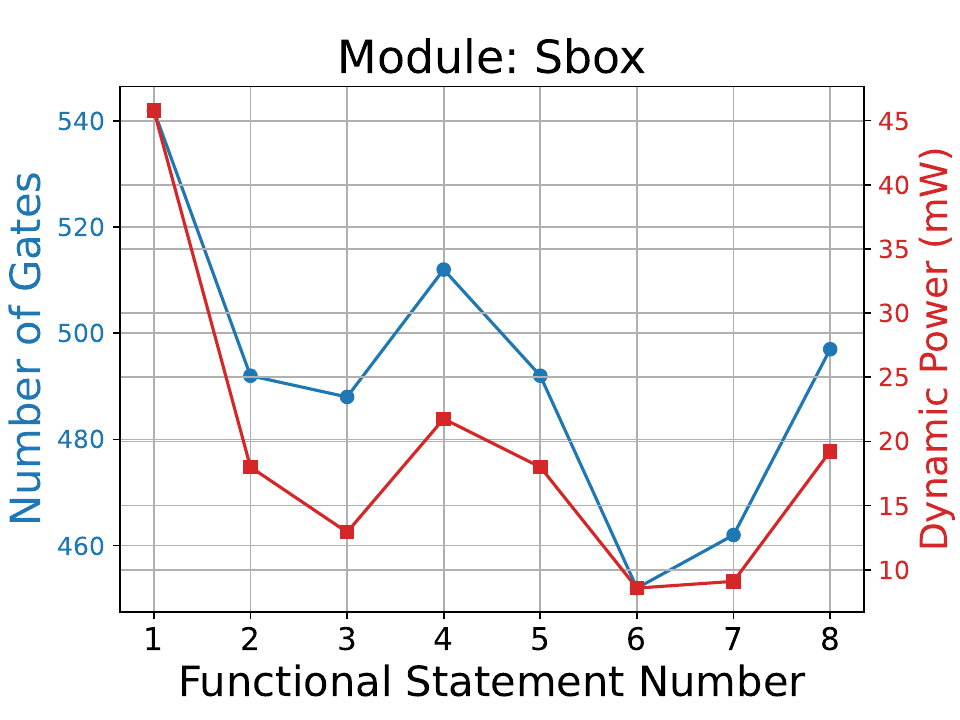}
         \caption{AES\_PPRM1}
         \label{fig:pprm1}
     \end{subfigure}
     \hfill
     \begin{subfigure}[b]{0.18\textwidth}
         \centering
         \includegraphics[width=\textwidth]{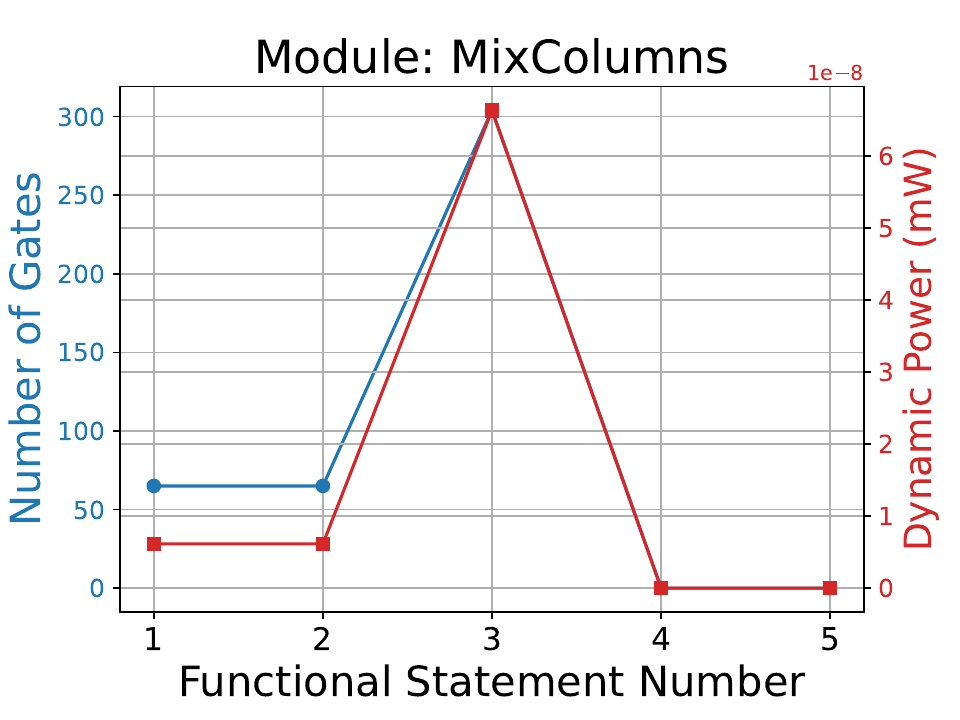}
         \caption{AES\_PPRM1}
         \label{fig:pprm12.}
     \end{subfigure}
     \hfill
     \begin{subfigure}[b]{0.18\textwidth}
         \centering
         \includegraphics[width=\textwidth]{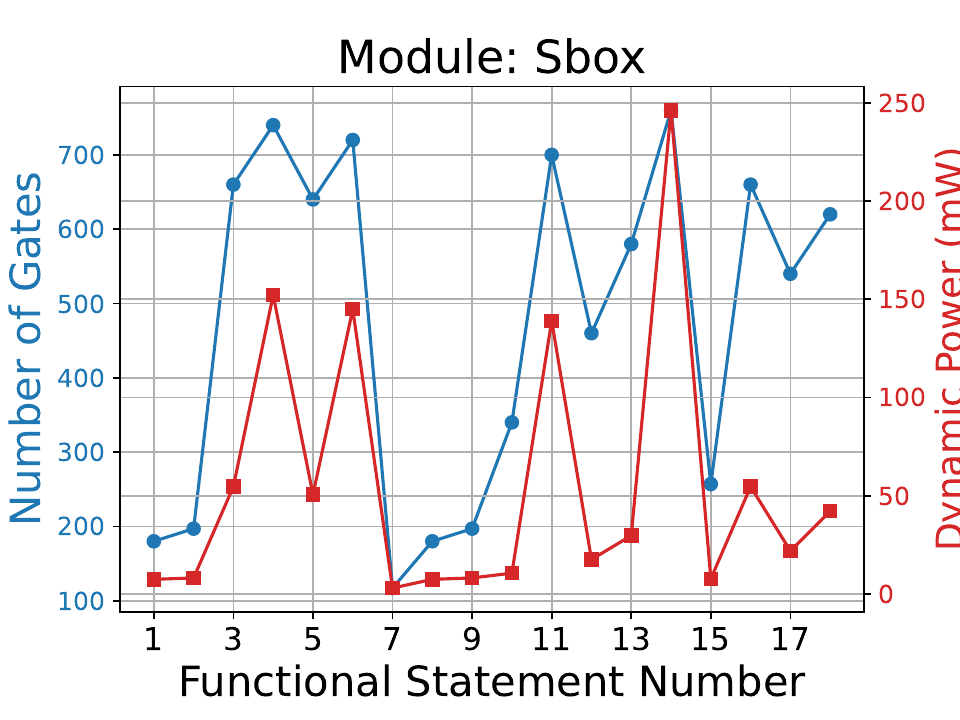}
         \caption{AES\_PPRM3}
         \label{fig:pprm31}
     \end{subfigure}
     \hfill
     \begin{subfigure}[b]{0.18\textwidth}
         \centering
         \includegraphics[width=\textwidth]{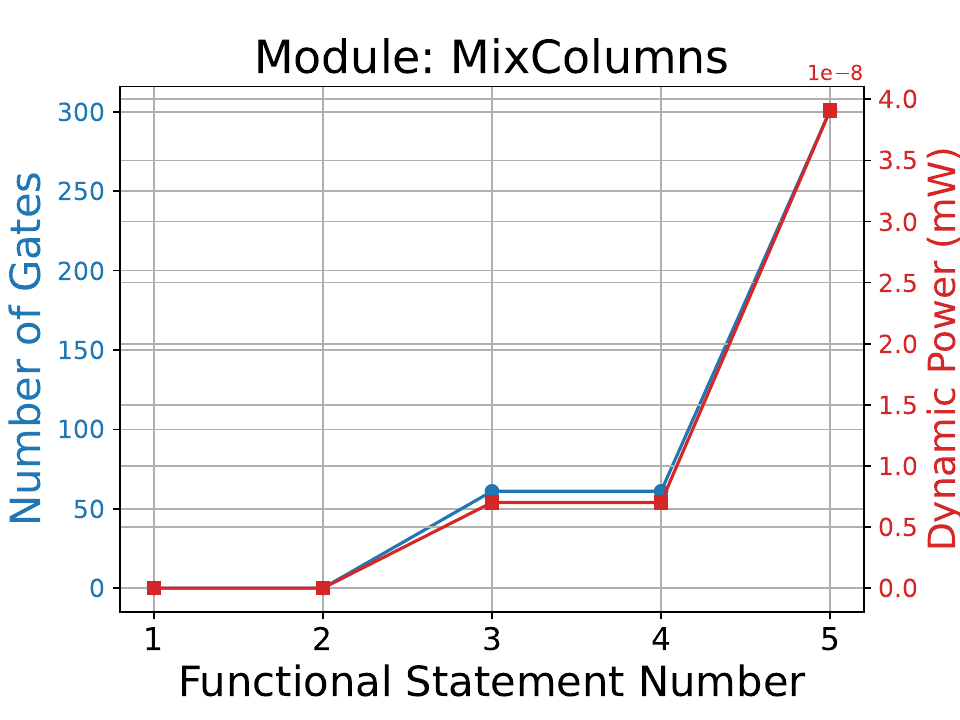}
         \caption{AES\_PPRM3}
         \label{fig:pprm32}
     \end{subfigure}
     \hfill
     \begin{subfigure}[b]{0.18\textwidth}
         \centering
         \includegraphics[width=\textwidth]{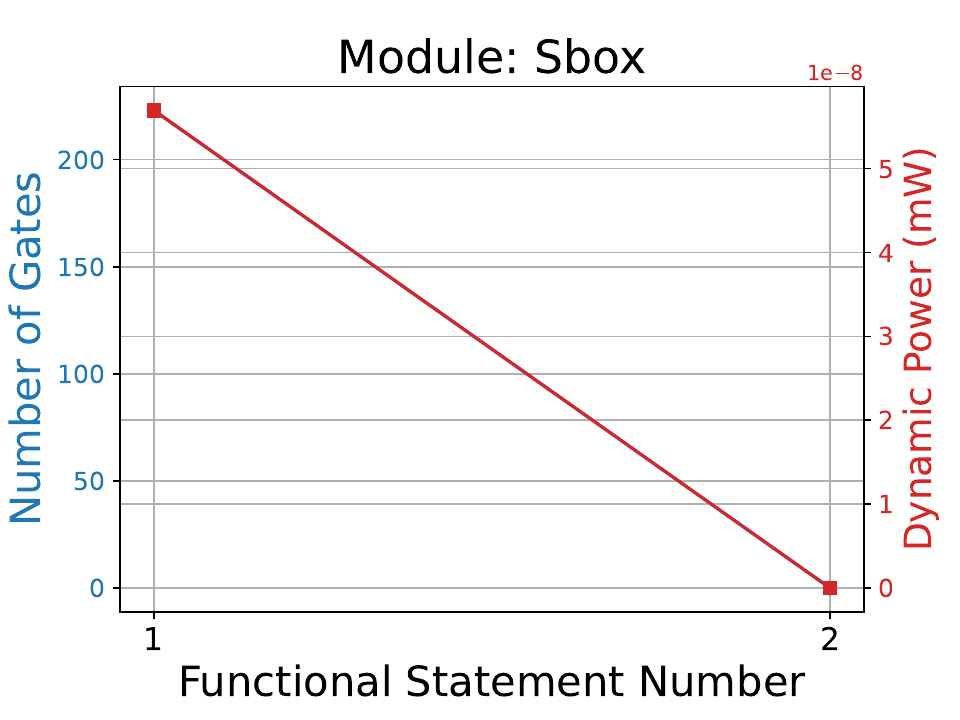}
         \caption{AES\_TBL}
         \label{fig:tbl1}
     \end{subfigure}
     \hfill
     \begin{subfigure}[b]{0.18\textwidth}
         \centering
         \includegraphics[width=\textwidth]{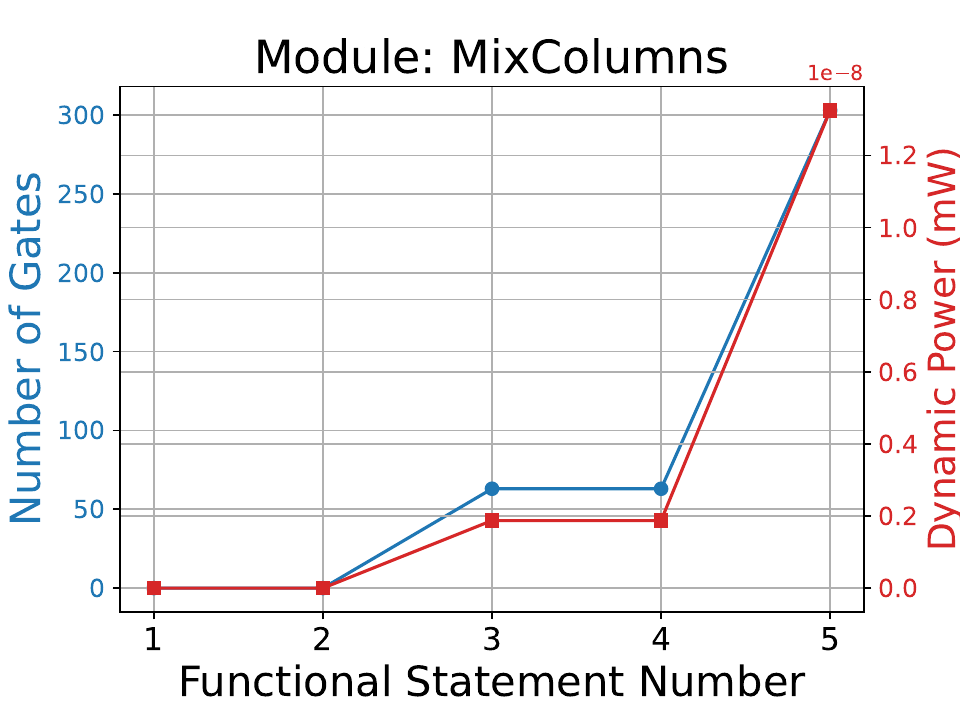}
         \caption{AES\_TBL}
         \label{fig:tbl2}
     \end{subfigure}
     \hfill
     \begin{subfigure}[b]{0.18\textwidth}
         \centering
         \includegraphics[width=\textwidth]{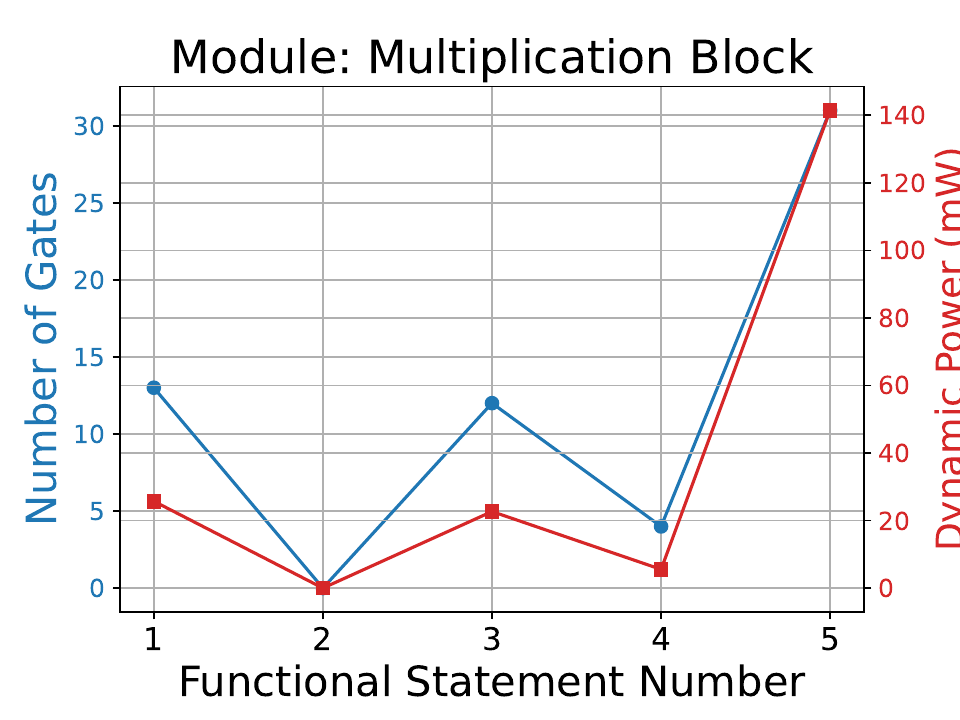}
         \caption{RSA1024\_RAM}
         \label{fig:rsa1}
     \end{subfigure}
     \hfill
     \begin{subfigure}[b]{0.18\textwidth}
         \centering
         \includegraphics[width=\textwidth]{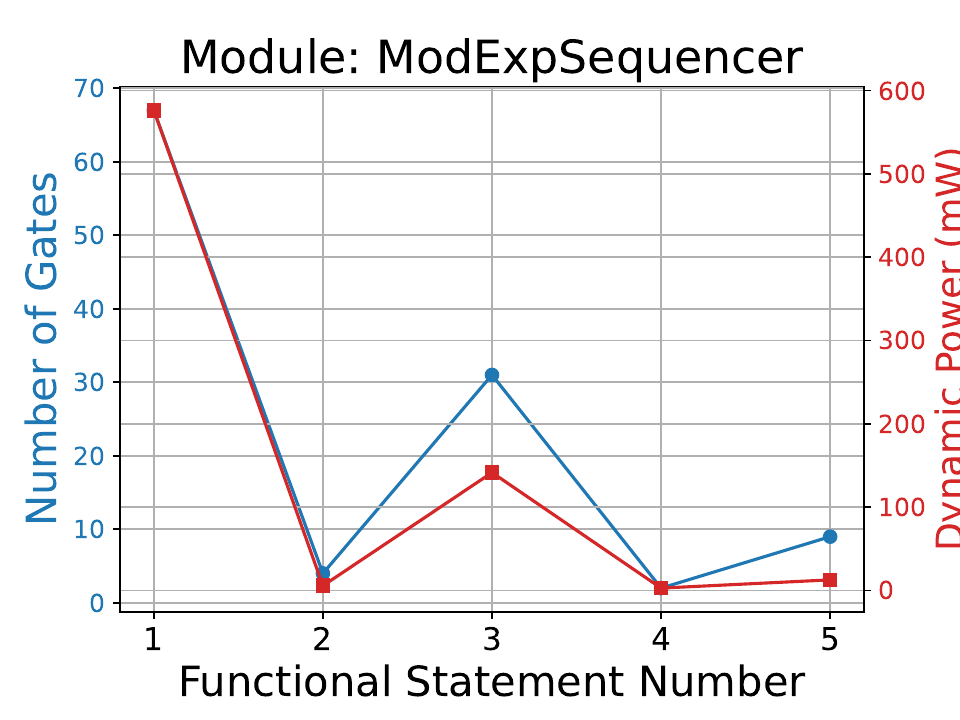}
         \caption{RSA1024\_RAM}
         \label{fig:rsa2}
     \end{subfigure}
     \hfill
     \begin{subfigure}[b]{0.18\textwidth}
        \centering
         \includegraphics[width=\textwidth]{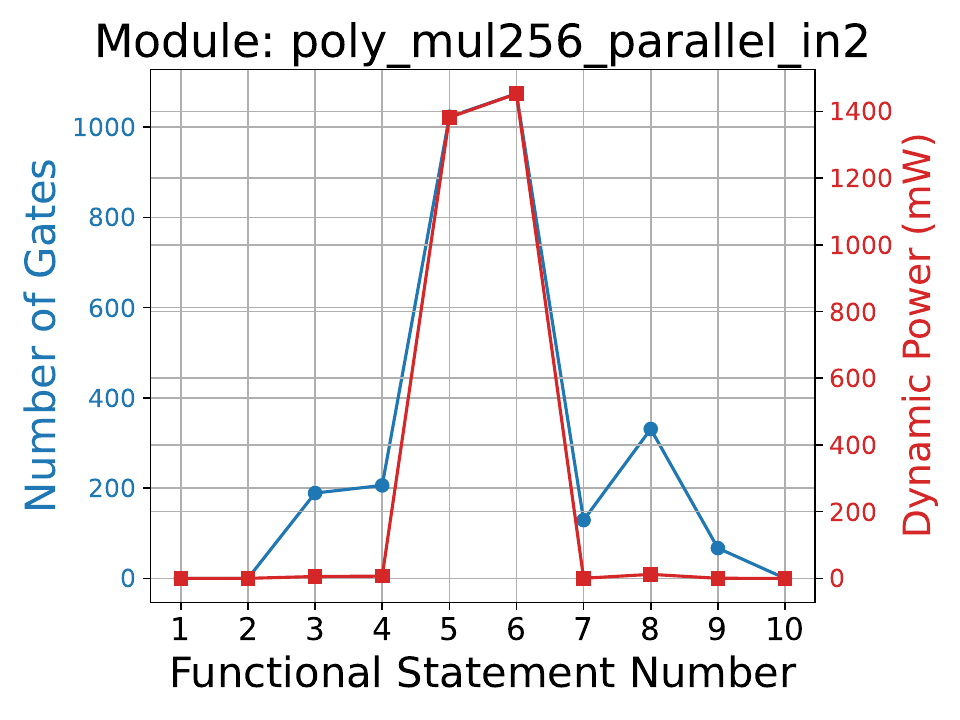}
         \caption{SABER}
         \label{fig:SABER1}
     \end{subfigure}
     \hfill
     \begin{subfigure}[b]{0.18\textwidth}
        \centering
         \includegraphics[width=\textwidth]{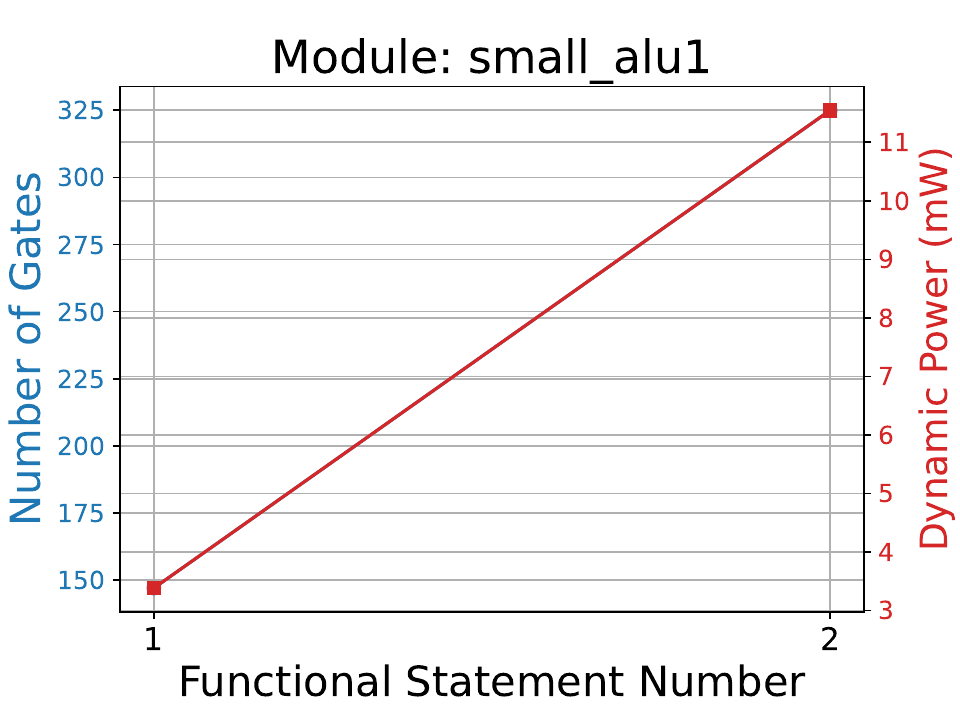}
         \caption{SABER}
         \label{fig:SABER2}
     \end{subfigure}
        \caption{These graphs depict the relationship between the dynamic power consumption and each location of the vulnerable module. The x-axis represents the `Functional Statement Number', which indicates the block number within the RTL source code of the vulnerable module. The y-axes show the number of gates in the synthesized netlist for each functional statement and their corresponding dynamic power consumption.}
        \label{fig:post}
\end{figure*}

In this section, our objective is to compute the dynamic power consumption of the RTL locations identified as vulnerable through our framework, SCAR. We intend to establish a comparative analysis by contrasting our prediction results with the quantitative dynamic power figures obtained post-synthesis, following the transformation of RTL designs into netlists.
For our post synthesis experiments, we utilize industry-scale tools from Synopsys: \textit{Synopsys VCS} for simulation and \textit{Synopsys Design Vision} for synthesis. Initially, we simulate our designs to ensure their functional correctness and performance. This is accomplished using designated test benches, which allow us to comprehensively verify the design's behavior under a variety of test cases. Following the successful simulation, we generate VCD files using \textit{Synopsys VCS}. 
The subsequent phase is design synthesis using \textit{Synopsys Design Vision}.
The tool transforms the RTL-level design description into gate-level representations, while accounting for critical factors such as timing, area, and power. The final result of this process is the generation of a netlist, a comprehensive representation of the entire design in terms of hardware primitives such as gates and flip-flops. Following the synthesis process, we analyze the number of gates synthesized in the netlist and the dynamic power consumption for each of the lines in the identified vulnerable modules. It is observed that the vulnerable lines predicted by our framework correspond to the highest number of gates and hence highest dynamic power amongst all other lines in the RTL design. 

Figure \ref{fig:post} provides a comprehensive overview of the post-synthesis reports for each of the benchmarks. Within these visual representations, the x-axis corresponds to the lines in the vulnerable modules of the RTL design, referred to as functional statements, since they compromise of various blocks for performing assignments, conditional statements such as if-else constructs, always blocks, case statements, and functions. Meanwhile, the y-axes represent both the number of gates present in the netlist and the associated dynamic power consumption. Please note that in the PRESENT benchmark, only the \textit{Sbox} module was vulnerable, which contains just one functional statement and therefore, it has not been visualized. Similarly, for the KYBER benchmark, the identified \textit{intmul} module also had only one functional statement. It was observed that the peak in dynamic power consumption occurs at the functional statements identified as `leaky' by our framework, as detailed in Table \ref{tab:summary}. It is noticeable that the functional statements which lead to highest dynamic power consumption, also have the highest number of gates in the netlist of the design. In the figures, we have represented the functional statement numbers, these numbers as well as their content is elaborated in Table \ref{tab:summary}. 

Our results align with existing research for accurately identifying vulnerable modules susceptible to PSC leakage in encryption algorithms \cite{pundir2022power}. While conventional approaches rely on complex power modeling for module-level identification, we have adopted a much simpler machine learning model. Furthermore, our approach also performs analysis at the RTL line-level, whose results have been verified with dynamic simulation on synthesized netlists. Through these results, it is evident that the locations pinpointed by SCAR in the design are associated with elevated dynamic power consumption. Notably, even when working with RTL designs, SCAR can produce highly accurate localizations early in the design phase, at par with those from post-synthesis netlist-based analysis.

\section{acknowledgment}
This research is partially supported by Technology Innovation Institute (TII), Abu Dhabi, UAE.
\section{Conclusion}\label{sec:Conclusion}


This paper proposes SCAR, a novel GNN-based approach to detect leaky modules and lines in encryption hardware RTL, which are vulnerable to PSC attacks. 
A notable limitation of existing post-silicon level PSC analysis approach lies in the potential impracticability of retrofitting security measures onto pre-existing devices, necessitating re-spins, and thus, increasing production cost. 
In contrast, SCAR, by leveraging the control data flow graph extracted from the encryption designs as well as an explainable GNN-based modeling, is able to capture the intricate relationships and dependencies between nodes (corresponding 
\balance
to design blocks in the RTL), enabling accurate identification of modules susceptible to PSC leakage in the pre-silicon RTL. 
Our proposed approach achieved up to 93.54\% accuracy, 100\% precision and 90.48\% recall, respectively, when evaluated on unseen AES implementations as well as unseen encryption algorithms, including Post-Quantum Cryptography algorithms. 
Furthermore, SCAR includes a robust source code analysis-based technique to pinpoint these vulnerabilities at a more granular RTL line-level. Subsequently, we employ an automated approach, leveraging a LLM, to effectively mitigate these identified vulnerabilities. We substantiate the accuracy of our findings by conducting a rigorous comparison with post-synthesis experiments. The proposed SCAR framework presents designers with an effective and proactive approach to mitigate the inherent risks associated with PSC attacks. Through the precise anticipation of vulnerable design blocks in advance, SCAR not only fortifies security measures but also yields substantial cost-savings. 



%

\bibliographystyle{IEEEtran}
\bibliography{references}
\end{document}